\documentclass[11pt,a4]{article}
\usepackage[latin1]{inputenc}
\usepackage{amsmath}
\usepackage{amsfonts}
\usepackage{amssymb}
\usepackage{graphicx}
\usepackage[toc,page]{appendix}
\usepackage{tocloft}
\usepackage{setspace}

\usepackage{bm}
\usepackage{float}
\usepackage{makeidx}
\usepackage{amsxtra}
\usepackage{tensor}
\usepackage{euscript}
\usepackage[margin=3cm]{geometry}

\def\ie{{\it i.e.}\ }
\def\eg{{\it e.g.}\ }

\def\be {\begin{equation}}
\def\ee {\end{equation}}
\def\bs#1\es{\begin{split}#1\end{split}}
\def\ba#1\ea{\begin{align}#1\end{align}}
\def\bg#1\eg{\begin{gathered}#1\end{gathered}}

\def\a{\alpha}
\def\b{\beta}
\def\c{\chi}
\def\d{\delta}

\def\e{\epsilon}
\def\ve{\varepsilon}
\def\f{\phi}
\def\vf{\varphi}
\def\F{\Phi}
\def\g{\gamma}

\def\h{\eta}
\def\k{\kappa}
\def\l{\lambda}
\def\L{\Lambda}
\def\m{\mu}
\def\n{\nu}
\def\o{\omega}
\def\p{\psi}

\def\r{\rho}
\def\s{\sigma}
\def\t{\tau}

\def\z{\zeta}

\def\pr{^\prime}
\def\pa{\partial}

\def\fr{\frac}

\def\blb{\bigg \lbrace}
\def\brb{\bigg \rbrace}
\def\bls{\bigg [}
\def\brs{\bigg ]}

\def\cD{\mathcal{D}}

\def\cP{\mathcal{P}}

\newcommand{\sq}[1]{\sqrt{#1}}

\def\uM{\underline M}
\def\uN{\underline N}

\def\uI{{\underline I}}
\def\uJ{{\underline{\phantom{I}}\!\!\!\! J}}

\def\ua{{\underline a}}
\def\ub{{\underline{\phantom{a}}\!\!\!\!b}}

\def\um{\underline \mu}
\def\un{\underline{\phantom{\mu}}\!\!\!\!\nu}

\def\um{\underline \mu}
\def\un{\underline{\phantom{\mu}}\!\!\!\!\nu}
\def\unn{\underline \nu}
\def\urho{\underline{\phantom{\mu}}\!\!\!\!\rho}
\def\usigma{\underline{\phantom{\mu}}\!\!\!\!\sigma}
\def\ulambda{\underline{\phantom{\mu}}\!\!\!\!\lambda}
\def\utau{\underline{\phantom{\mu}}\!\!\!\!\tau}

\def\uzero{\underline{\phantom{\mu}}\!\!\!\! 0}
\def\uone{{\underline{\phantom{\mu}}\!\!\!\! 1}}
\def\utwo{\underline{\phantom{\mu}}\!\!\!\! 2}
\def\uthree{\underline{\phantom{\mu}}\!\!\!\! 3}
\def\ufour{\underline{\phantom{\mu}}\!\!\!\! 4}
\def\ufive{\underline{\phantom{\mu}}\!\!\!\! 5}
\def\useven{\underline{\phantom{\mu}}\!\!\!\! 7}
\def\usevenn{{\underline 7}}

\def\nn{\nonumber}
\def\bea{\begin{eqnarray}}
\def\eea{\end{eqnarray}}
\def\ft#1#2{{\textstyle{{\scriptstyle #1}\over {\scriptstyle #2}}}}

\newcommand{\w}[1]{\\[0.#1cm]}

\def\tr{{\rm tr}}
\def\Tr{\rm Tr}

\def\tw{\text{\tiny{(2)}}}
\def\th{\text{\tiny{(3)}}}
\def\fo{\text{\tiny{(4)}}}
\def\we{\wedge}

\def\ztwo{\mathbb{Z}_2}

\def\eq#1{(\ref{#1})}
\def\ev{\big|_{\partial M}}

\numberwithin{equation}{section}
\numberwithin{figure}{section}

\newcommand{\hoch}[1]{$\, ^{#1}$}

\newcommand{\auth}{\large
T.G. Pugh,\footnote{email: thomas.pugh08@imperial.ac.uk} E. Sezgin\footnote{email: sezgin@physics.tamu.edu}  and K.S. Stelle\footnote{email: k.stelle@imperial.ac.uk}}

\newcommand{\tamphys}{\it George and Cynthia Woods Mitchell  Institute
for Fundamental Physics and Astronomy,\\
Texas A\&M University, College Station, TX 77843, USA}

\newcommand{\imperial}{\it The Blackett Laboratory,
Imperial College, Prince Consort
Road, London SW7 2BZ, UK}

\newcommand{\golm}
{\it AEI, Max Planck Institut f\"ur Gravitationsphysik,
Am M\"{u}hlenberg 1, D-14476 Potsdam, Germany}

\begin{document}

\renewcommand{\thefootnote}{\fnsymbol{footnote}}

\begin{flushright}
\hfill{
MIFPA-10-30}\\
\hfill{Imperial/TP/10/KSS/01}\\
\hfill{AEI-2010-123}
\end{flushright}

\vspace{20pt}

\begin{center}

\begin{doublespace}

{\Large\bf $D=7\slash D=6$ Heterotic Supergravity\\ with Gauged R-Symmetry}

\end{doublespace}

\vspace{10pt}

\auth

\vspace{10pt}

\hoch{\ast\ \ddagger}{\imperial}\\
\vskip 1 em
\hoch{\dagger}{\tamphys}\\
\vskip 1 em
\hoch{\ddagger}{\golm}

\vspace{20pt}

\underline{ABSTRACT}
\end{center}

We construct a family of chiral anomaly-free supergravity theories in $D=6$ starting from $D=7$ supergravity with a gauged noncompact R-symmetry, employing a Ho\v{r}ava-Witten bulk-plus-boundary construction. The gauged noncompact R-symmetry yields a positive (de Sitter sign) $D=6$ scalar field potential. Classical anomaly inflow which is needed to cancel boundary-field loop anomalies requires careful consideration of the gravitational, gauge, mixed and local supersymmetry anomalies. Coupling of boundary hypermultiplets requires care with the $\mathrm{Sp}(1)$ gauge connection required to obtain quaternionic K\"ahler target manifolds in $D=6$. This class of gauged R-symmetry models may be of use as starting points for further compactifications to $D=4$ that take advantage of the positive scalar potential, such as those proposed in the scenario of supersymmetry in large extra dimensions.

\vspace{40pt}

\renewcommand{\thefootnote}{\arabic{footnote}}
\setcounter{footnote}{0}

\pagebreak

\tableofcontents

\newpage

\section{Introduction}

Anomaly-free chiral $N=(1,0)$ gauged supergravities in $D=6$ \cite{Nishino:1986dc,Nishino:1997ff,Ferrara:1997gh,Riccioni:2001bg,RandjbarDaemi:1985wc,Avramis:2005qt,Avramis:2005hc} have intriguing possible phenomenological applications, in particular for scenarios involving supersymmetry in large extra dimensions \cite{Aghababaie:2003wz,Burgess:2007ui}. A key challenge with such supergravity models has been to embed them in string or M-theory while also ensuring the absence of quantum gravitational or gauge anomalies. One way to generate anomaly-free chiral models is the Ho\v{r}ava-Witten mechanism \cite{Horava:1996ma,Horava:1995qa}, which involves compactification on a line interval  while at the same time supposing that matter fields appear on the end-walls of the interval in such a combination as to cancel the quantum anomalies. The basic Ho\v{r}ava-Witten scenario involves a stage of Kaluza-Klein reduction on $S^1/\ztwo$ followed by a search for anomaly-cancelling matter combinations with which to populate the bounding walls. In order to obtain an $N=(1,0), D=6$ theory with gauged $U(1)$ R-symmetry in this way, one would need to begin this stage of reduction with an appropriate $D=7$ theory. For this purpose, we shall use the construction of Ref.\ \cite{Cvetic:2003xr} which achieved $D=6$ R-symmetry starting from $N=1, D=10$ supergravity and reducing on the noncompact space $H(2,2)$, which is endowed with a Euclidean signature metric of cohomogeneity one. This produces a theory containing minimal $D=7$ supergravity coupled to Super Yang-Mills with an $SO(2,2)$ noncompact gauge group. The noncompact nature of this $D=7$ gauge group is essential for allowing subsequent truncation to a chiral $D=6$ theory that retains an R-symmetry gauging of the sort found in Ref.\ \cite{Salam:1984cj}.

Reduction on a noncompact space obviously raises a number of important issues which would need to be addressed before such a construction could be considered physically reasonable. We will comment on this problem, but this issue will not be our main focus here. Rather, we will focus on another major problem arising with chiral $D=6$ gauged supergravity models: ensuring the absence of mixed gravitational, supersymmetry and gauge anomalies. The anomaly analysis of Ref.\ \cite{Horava:1996ma,Horava:1995qa} for the reduction of $D=11$ M-theory on $S^1/\ztwo$ yielded $E_8$ Super-Yang-Mills matter multiplets on each of the two $D=10$ bounding walls. A similar analysis involving the reduction of the $D=7$ theory obtained in \cite{Cvetic:2003xr} on $S^1/\ztwo$ down to $D=6$ will be our main focus in the present paper. $SU(2)$ gauged half-maximal $D=7$ supergravity, and its coupling to vector multiplets have been studied on a manifold with boundaries in Refs \cite{Gherghetta:2002nq,Avramis:2004cn}. There are important differences in the models considered in these papers and the ones we study in this paper, the most important one being that, unlike in \cite{Gherghetta:2002nq,Avramis:2004cn}, we here maintain R-symmetry gauging on the boundary. As  mentioned above, starting from a noncompact gauge theory in $D=7$ is essential for this to work. Furthermore, we will study the couplings of the scalar fields surviving the $\ztwo$ projection on the boundary, and will determine the complete set of boundary conditions needed for closure under supersymmetry.

In Section \ref{sectwo} we will review the $N=1, D=7$ gauged supergravity which will describe our bulk theory \cite{BergKohSez}. This can be obtained starting from $N=1, D=10$ supergravity reduced on $H(2,2)$ as in \cite{Cvetic:2003xr}. Then, in Section \ref{orbifold} we will go on to consider this theory on an $S^1/\ztwo$ orbifold and we will demonstrate the necessity of appropriate Gibbons-Hawking-York terms. After this, we will continue on in Section \ref{7to6} to consider a dimensional reduction of the $D=7$ bulk theory to $D=6$ by taking a limit of vanishing orbifold width. This will be necessary to prepare the appropriate variables for subsequent bulk-boundary coupling.

The coupling of $D=6$ supersymmetric boundary-localised matter to the $D=7$ bulk theory involves some delicate steps. In Sections \ref{BCs} and \ref{BoundaryAction}, we will concentrate on the coupling of boundary $D=6$ vector multiplets to the $D=7$ bulk fields. This involves, firstly, a careful consideration of how the boundary conditions for the bulk fields need to be modified in the presence of the boundary fields, as discussed in Section \ref{BCs}.

Since the {\it raison d'\^etre} of the boundary fields is to provoke a ``classical'' anomalous gauge variation which can be used to compensate for quantum anomalies occurring via quantum loops on the $D=6$ boundaries, one expects the bulk-plus-boundary field construction to produce a non-vanishing variation under gauge symmetries. However, since the closure of the supersymmetry algebra generates gauge transformations, one finds that the classical gauge anomalies are accompanied by classical supersymmetry anomalies as well. Accordingly, one cannot carry out the construction of the bulk-plus-boundary system while requiring exact supersymmetry invariance. Instead, one must be guided by the necessity of ensuring that the total variation of the bulk-plus-boundary system satisfies the Wess-Zumino consistency conditions, in order to have the structure necessary to cancel anomalies that will arise from boundary-field quantum loops. This is discussed in Section \ref{BoundaryAction}.

In Section \ref{boundaryhypers}, we will consider the coupling of boundary-localised hypermultiplets. This proceeds in a similar way to the coupling of the boundary vector multiplets. However, as there is no bosonic anomaly associated to the hypermultiplets, there will be no corresponding supersymmetry anomaly. The coupling of hypermultiplets is complicated by the fact that the scalars of the bulk and boundary sectors are required to combine to form a quaternionic K\"{a}hler manifold (QKM). We will demonstrate that this imposes a constraint on the $Sp(1)$ connection of the boundary sector which sets it equal to the $Sp(1)$ connection of the bulk.

The models we construct in Sections \ref{BCs} and \ref{BoundaryAction} will be Wess-Zumino consistent, but will not yet provide the full set of classical anomalies that are needed to cancel all the quantum anomalies. In Section \ref{extensions}, we will consider extensions of the present model that can give rise to the remaining cancellations. We will consider the supersymmetric extension of the the bulk model Chern-Simons terms, focusing particularly on a topological mass term. As well as examining alternative boundary conditions, we finally will look at the coupling of boundary-localised tensor multiplets.

In Section \ref{anomalies}, we will consider an explicit example of an anomaly-cancelling system. To do this, we will calculate the anomaly polynomial produced by one-loop quantum effects. We will then show how the Wess-Zumino consistent classical anomalies constructed so far can be arranged so as to cancel these quantum anomalies.

In Appendix A, we examine the limit of coincident boundaries when the boundaries are populated with vector multiplets and will show the emergence of gauged supergravity in $D=6$. In Appendix B, we will provide the bulk-plus-boundary construction with a supersymmetric set of boundary conditions in an equivalent formulation in which the bulk 3-form potential is dualised to a 2-form potential.

\section{$D=7$ 3-Form Supergravity}\label{sectwo}

Seven dimensional $N=1$ supergravity in the absence of boundaries has been well studied, and the action of the supergravity multiplet coupled to $n$ vector multiplets is known \cite{BergKohSez}. The fields in this action form a reducible multiplet with field content,
\begin{equation}
(\hat{e}_M^{\uN}, \hat{A}_{MNR}, \hat\sigma, \hat{A}_M^{\hat{I}}, \phi_\alpha ,  \hat{\psi}_M^A , \hat\chi^A, \hat{\lambda}^{\hat{r} A})\ ,
\end{equation}
where $M = 0,\ldots,5,7$ is the world index, which is raised and lowered with the metric $\hat{g}_{MN}$ and $\uN = \underline{0},\ldots,\underline{5},\underline{7}$  is the tangent-space lorentz index, which is raised and lowered with the metric $\eta_{\uM\uN} = \mathrm{diag}(- + \dots +) $.

The scalars $\phi_\alpha$ with $\alpha = 1,2,...,3n $ parametrise a coset,
\begin{equation}
\frac{SO(n,3)}{SO(n) \times SO(3)}\ ,
\label{coset}
\end{equation}
for which we can form the representative elements $L_{\hat{I}}^{i}$ and $L_{\hat{I}}^{\hat{r} }$, where $\hat{I} = 1,\ldots,n+3$ is an $SO(n,3)$ index, which is raised and lowered with the $SO(n,3)$ invariant metric $\eta_{\hat{I} \hat{J}} = \mathrm{diag} ( - - - + \ldots +)$.  $i= 1,\ldots,3$ is an $SO(3)$ index and $\hat{r} = 1,\ldots,n$ is an $SO(n)$ index; these are raised and lowered with the Kronecker deltas $\delta_{ij}$ and $\delta_{\hat{r} \hat{s}}$ respectively. The coset representatives satisfy the relations
\begin{equation}
-L_{\hat{I}}^i L_{\hat{J}}^i + L_{\hat{I}}^{\hat{r}} L_{\hat{J}}^{\hat{r}} = \eta_{\hat{I}\hat{J}}\ ,
\end{equation}
\begin{equation}
L_{\hat{I}}^i L_j^{\hat{I}} = -\delta_j^i\ , \quad L_{\hat{I}}^{\hat{r}} L^{\hat{I}}_{\hat{s}} = \delta_{\hat{s}}^{\hat{r}}\ , \quad L_{\hat{I}}^i L_{\hat{r}}^{\hat{I}} = 0\  .
\end{equation}
The spinors are symplectic Majorana and carry an $Sp(1)$ doublet index $A = 1,2$ which is raised and lowered with the metric $\e_{AB}$\footnote{Our conventions are: $\psi^A=\e^{AB}\psi_B\ , \psi_A=\psi^B \e_{BA}$ and $\e_{AB}\e^{BC}=-\delta_A^C$.}. The $Sp(1)$ indices will often be suppressed, as in $ \bar\chi \sigma^i  \epsilon = \bar\chi^A {{\sigma^i}_A}^B \epsilon_B$.

The action for these fields, up to terms quadratic in fermions, is given by
\be
\bs
S_{SG} =& \fr1{2\kappa^2} \int  d^7x \hat{e} \bigg \lbrace \frac12 \hat{R}  - \frac1{4 g^2} e^{\hat{\sigma}}
\hat{F}_{MN}^i \hat{F}^{MN i } - \frac1{4g^2}e^{\hat{\sigma}} \hat{F}_{MN}^{\hat{r}} \hat{F}^{MN \hat{r}}\\
& - \frac{1}{48}  e^{-2 \hat{\sigma}} \hat{F}_{MNRS} \hat{F}^{MNRS}  - \frac1{24 \sqrt2 g^2} \hat\varepsilon^{MNRSTUV} \hat{A}_{MNR} \hat{F}_{ST}^{\hat r} {\hat F}_{UV}^{\hat r} \\
& - \frac58 \partial_M \hat{\sigma} \partial^M \hat{\sigma} -\frac12 \hat{P}_M^{i \hat{r}} \hat{P}^{M i \hat{r}} - \frac14 g^2 e^{- \hat{\sigma} }  \left( C^{i \hat{r}} C^{i \hat{r}} - \frac19 C^2 \right)\\
&-\frac{i}{2} \hat{\bar{\psi}}_M \hat{\gamma}^{MNR} \hat{D}_N \hat{\psi}_R - \frac{5i}{2}  \hat{\bar{ \chi}} \hat{\gamma}^M \hat{D}_M \hat\chi - \frac{i}{2g^2} \hat{\bar{\lambda}}^{\hat{r}} \hat{\gamma}^M \hat{D}_M \hat{\lambda}_{\hat{r}} \\
&- \frac{5i}{4} \hat{\bar{\chi}} \hat{\gamma}^M \hat{\gamma}^N \hat{\psi}_M \partial_N \hat{\sigma} - \frac1{2g} \hat{\bar{\lambda}}^{\hat{r}} \sigma^i \hat{\gamma}^M \hat{\gamma}^N \hat{\psi}_M \hat{P}_N^{i \hat{r}}\\
\es
\ee
\be
\bs
& + \frac{i}{96 \sqrt{2}} e^{-\hat{\sigma}} \hat{F}_{MNRS} \bigg ( \hat{\bar\psi}_{[L} \hat\gamma^L \hat\gamma^{MNRS} \hat\gamma^T \hat\psi_{T]} + 4 \hat{\bar\psi}_L \hat{\gamma}^{MNRS} \hat\gamma^L \hat\chi \\
&- 3 \hat{\bar\chi} \hat\gamma^{MNRS} \hat\chi + \frac1{g^2} \hat{\bar\lambda}^{\hat{r}} \hat\gamma^{MNRS} \hat{\lambda}^{\hat{r}} \bigg)    + \frac1{8g} e^{\frac{\hat\sigma}2} \hat{F}_{MN}^i \bigg( \hat{\bar\psi}_{[L} \sigma^i  \hat\gamma^{L} \hat\gamma^{MN} \hat{\gamma}^T \hat\psi_{T]} \\
&- 2 \hat{\bar\psi}_L \sigma^i \hat\gamma^{MN} \hat\gamma^L \hat\chi + 3  \hat{\bar\chi} \sigma^i \hat\gamma^{MN} \hat{\chi} - \frac1{g^2} \hat{\bar\lambda}^{\hat{r}} \sigma^i \hat\gamma^{MN} \hat\lambda^{\hat{r}} \bigg) \\
& - \frac{i}{4g^2} e^{\frac{\hat{\sigma}}{2}} \hat{F}_{MN}^{\hat{r}} \bigg( \hat{\bar\psi}_L \hat\gamma^{MN} \hat\gamma^{L} \hat\lambda^{\hat{r}} + 2 \hat{\bar\chi} \hat\gamma^{MN} \hat\lambda^{\hat{r}} \bigg) \\
&- \frac{i\sqrt{2}}{24} g e^{-\frac{\hat\sigma}{2}} C \bigg( \hat{\bar\psi}_M \hat\gamma^{MN} \hat\psi_N +2 \hat{\bar\psi}_M \hat\gamma^M \hat\chi + 3 \hat{\bar\chi} \hat{\chi} - \frac1{g^2} \hat{\bar\lambda}^{\hat{r}} \hat{\lambda}^{\hat{r}} \bigg)\\
& + \frac{1}{2 \sqrt{2}} e^{-\frac{\hat{\sigma}}{2}} C^{i \hat{r}} \bigg( \hat{\bar\psi}_M \sigma^i \hat\gamma^M \hat\lambda^{\hat{r}} - 2 \hat{\bar\chi} \sigma^i \hat\lambda^{\hat{r}} \bigg) + \frac1{2g} e^{-\frac{\hat\sigma}{2}} C^{\hat{r} \hat{s} i} \hat{\bar\lambda}^{\hat{r}} \sigma^i \hat\lambda^{\hat{s}}  \bigg \rbrace\ ,
\label{7DAction}
\es
\ee
where  $\hat{F}_{MNRS} = 4 \partial_{[M} A_{NRS]}$ is the field strength, invariant under tensor gauge transformations $\delta {\hat A}_{MNR}=3\partial_{[M} {\hat\lambda}_{NR]}$;  $\hat{\varepsilon}^{\uzero\uone\utwo\uthree\ufour\ufive\useven}= 1$. Furthermore,
\be
\bs
\hat{F}^{\hat{I}}_{MN} &= 2 \partial_{[M} \hat{A}_{N]}^{\hat{I}} + {f_{\hat{J} \hat{K}}}^{\hat{I}} \hat{A}_M^{\hat{J}} \hat{A}_N^{\hat{K}}\ ,
\w2
\hat{F}_{MN}^i &= \hat{F}_{MN}^{\hat{I}} L_{\hat{I}}^i, \quad \hat{F}_{MN}^{\hat{r}} = \hat{F}_{MN}^{\hat{I}} L_{\hat{I}}^{\hat{r}}\ ,
\w2
\hat\omega_{MNR}^0 (\hat{A}^{\hat I }) &= \hat{F}_{[MN}^{\hat I} \hat{A}_{R]}^{\hat I} - \frac13 f_{\hat I \hat J \hat K} \hat{A}_M^{\hat I} \hat{A}_N^{\hat J} \hat{A}_R^{\hat K}\ .
\label{fs}
\es
\ee

\ba
\hat{D}_M &= \hat{\partial}_M + \frac{1}{4} \hat{\omega}_{\mu \um \un } \gamma^{\um\un } + \frac{1}{2 \sqrt{2}} \hat{Q}_M^i \sigma^i\ , & \hat{Q}_M^i &= \frac{i}{\sqrt{2}} \epsilon^{ijk} \hat{Q}_M^{jk}\ ,
\nn\w2
\hat{Q}_M^{i j} &= L^{\hat{I} j} \left( \delta_{\hat{I}}^{\hat{K}} \partial_M +  {f_{\hat{I} \hat{J}}}^{\hat{K}} \hat{A}_M^{\hat{J}} \right) L_{\hat{K}}^i\ , &
\hat{Q}_M^{\hat{r} \hat{s}} &= L^{\hat{I} \hat{r}} \left( \delta_{\hat{I}}^{\hat{K}} \partial_M +  {f_{\hat{I} \hat{J}}}^{\hat{K}} \hat{A}_M^{\hat{J}} \right) L_{\hat{K}}^{\hat{s}}\ ,
\nn\w2
\hat{P}_M^{i \hat{r}} &= L^{\hat{I} \hat{r}} \left( \delta_{\hat{I}}^{\hat{K}} \partial_M +  {f_{\hat{I} \hat{J}}}^{\hat{K}} \hat{A}_M^{\hat{J}} \right) L_{\hat{K}}^i\ , &
C &= - \frac{1}{\sqrt{2}} {f_{\hat{I} \hat{J}\hat{K}}}  L^{\hat{I} i }  L^{\hat{J} j } L^{\hat{K} k } \epsilon^{ijk}\ ,
\nn\w2
C^{i \hat{r}} &= \frac{1}{\sqrt{2}} {f_{\hat{I} \hat{J}\hat{K}}}  L^{\hat{I} j }  L^{\hat{J} k } L^{\hat{K} \hat{r} } \epsilon^{ijk}\ , &
C^{\hat{r} \hat{s} i } &= {f_{\hat{I} \hat{J}\hat{K}}}  L^{\hat{I} \hat{r} }  L^{\hat{J} \hat{s} } L^{\hat{K} i }\ ,
\ea
and $\hat{R} $ is the curvature defined with respect to the torsion-free Levi-Civita connection.  The vectors $\hat{A}_M^{\hat{J}}$ gauge a group $K \subset SO(n,3)$  with $n+3$ generators. Possible gauge groups are discussed in \cite{Bergshoeff:2005pq}. Of special interest are certain non-compact gauge groups which allow an R-Symmetry gauging upon dimensional reduction to $D=6$ followed by chiral truncation. We shall make restrictions to such gaugings in Section \ref{orbifold}, but for now we will leave the construction general.

The action is invariant under the following local supersymmetry transformations,
\begin{equation}
\begin{split}
\delta \hat{e}_M^{\uM} &= i \hat{\bar\epsilon} \gamma^{\uM} \hat\psi_M\ ,\\
\delta \hat{\psi}_M &=  2 \hat{D}_M \hat\epsilon - \frac{1}{240 \sqrt{2}} e^{- \hat\sigma} \hat{F}_{RSLT} \left( \hat\gamma_M \hat\gamma^{RSLT} + 5 \hat\gamma^{RSLT} \hat\gamma_M \right) \hat\epsilon \\
&- \frac{i}{20g} e^{\frac{\hat\sigma}{2}} \hat{F}_{RS}^i \sigma^i \left( 3 \hat\gamma_M \hat\gamma^{RS} - 5 \hat\gamma^{RS} \hat\gamma_M \right) \hat\epsilon - \frac{\sqrt{2}}{30} g e^{-\frac{\hat\sigma}{2}} C \hat\gamma_M \hat{\epsilon}\ ,\\
\delta \hat\chi &= - \frac12 \hat\gamma^M \hat{\partial}_M \hat\sigma  \hat\epsilon- \frac{1}{60 \sqrt{2}} e^{- \hat\sigma} \hat{F}_{MNRS} \hat\gamma^{MNRS} \hat\epsilon \\
&  - \frac{i}{10g} e^{\frac{\hat\sigma}{2}} \hat{F}_{MN}^i \sigma^i \hat\gamma^{MN} \hat{\epsilon} + \frac{\sqrt{2}}{30}  g e^{-\frac{\hat\sigma}{2}} C \hat\epsilon\ ,\\
\delta \hat{A}_{MNR} &= \frac{3i}{\sqrt{2}} e^{\hat{\sigma}} \hat{\bar\epsilon} \hat\gamma_{[MN}\hat\psi_{R]}  - i \sqrt 2 e^{\hat{\sigma}} \hat{\bar\epsilon} \gamma_{MNR} \hat\chi\ ,\\
\delta \hat{A}_M^{\hat{I}} &= - g e^{-\frac{\hat\sigma}{2}} \left( \hat{\bar\epsilon} \sigma^i \hat\psi_M + \hat{\bar\epsilon} \sigma^i \hat\gamma_{M} \hat\chi \right) L^{\hat{I} i} + i e^{-\frac{\hat\sigma}{2}} \hat{\bar{\epsilon}} \hat\gamma_M \hat\lambda^{\hat{r}} L^{\hat{I} \hat{r}}, \\
\delta \hat\sigma &= - 2 i \hat{\bar\epsilon} \hat\chi\ ,\\
\delta L_{\hat{I}}^i &= \frac{1}{g}\, \hat{\bar\epsilon} \sigma^i \hat\lambda^{\hat{r}} L_{\hat{I}}^{\hat{r}}\ ,\\
\delta L_{\hat{I}}^{\hat{r}} &= \frac1{g}\, \hat{\bar\epsilon} \sigma^i \hat{\lambda}^{\hat{r}} L_{\hat{I}}^i \ ,\\
\delta \hat\lambda^{\hat{r}} &= -\frac12 e^{\frac{\hat\sigma}{2}} \hat{F}_{MN}^{\hat{r}} \hat\gamma^{MN} \hat\epsilon + i g \hat\gamma^M \hat{P}_M^{i \hat{r}} \sigma^i \hat\epsilon - \frac{i}{\sqrt{2}} g e^{-\frac{\hat\sigma}{2}} C^{i \hat{r}} \sigma^i \hat\epsilon\ .
\end{split}
\label{7DSUSY}
\end{equation}

\section{The Model on an $S^1/{\ztwo}$ Orbifold}\label{orbifold}

The action has a $\ztwo$ parity symmetry under which $x^7 \rightarrow -x^7$, and the following fields have even parity:
\begin{equation}
( \hat{e}_{\mu}{}^{\unn}, \hat{e}_{7}{}^{\usevenn} , \hat{A}_{ \mu \nu 7 }, \hat\sigma, \hat{A}_\mu^{I^\prime}, \hat{A}_7^I, \phi^{ri}, \hat\psi_{\um +}, \hat\psi_{\usevenn -}, \hat\chi_-, \hat\lambda^r_-, \hat\lambda^{r^\prime}_+)\ ,
\label{even}
\end{equation}
whilst the odd-parity fields are
\begin{equation}
(\hat{e}_{\mu}{}^{\usevenn},\hat{e}_{7}{}^{\unn}, \hat{A}_{\mu \nu \rho}, \hat{A}_\mu^I, A_7^{I^\prime},\phi^{r'i}, \hat\psi_{\um -}, \hat\psi_{\usevenn +}, \hat\chi_+, \hat\lambda^r_+, \hat\lambda^{r^\prime}_-)\ ,
\label{odd}
\end{equation}
where  the scalars $(\phi^{ri}, \phi^{r'r})$ parametrize the coset \eq{coset}. The supersymmetry transformation rules are consistent with these parity assignments provided that $\e_+$ has even parity and $\e_-$ has odd parity.  In the definitions \eqref{even} and \eqref{odd}, we have split up the index $M$ into the $7$ direction and the directions normal to it, which are labelled by $\mu = 0,\ldots,5$. We have also defined a chiral projection operator $P_\pm = \frac{1}{2} \left( 1 \pm \gamma^{\usevenn} \right)$, which projects onto chiral spinors in the standard way, \ie $\chi_\pm = P_\pm \chi$. The $\hat{r}$ and $\hat{I}$ indices  have also been split as
\ba
\hat{I} &= \lbrace I\ , I^\prime \rbrace, & I &= 1,...,p+3\ , & I^\prime &= p+4,..., n+3
\nn\w2
\hat{r} &= \lbrace r\ , r^\prime \rbrace, & r &= 1,..., p\ , & r^\prime &= p+1, ..., n
\label{indexsplit}
\ea
where $0 \leq p \leq n $. Next, we observe that the requirement that the Yang-Mills field strength \eq{fs} have a definite parity imposes the conditions
\begin{equation}
{f_{IJ}}^K = {f_{I^\prime J^\prime}}^K = 0 \ .
\end{equation}
The possible groups K which posses this property and which reduce to give a gauged supergravity in 6 dimensions are $SO(3,1)$, $SO(2,1)$ and $SO(2,2)$ \cite{Bergshoeff:2005pq}. Since the action is invariant under a $\ztwo$ symmetry, we can formulate the action integral on a manifold $M \times I$, where $M$ is an arbitrary $D=6$ spacetime and $I=S_1/\ztwo$ is an interval with boundaries ($ \pa M$) at $x^7=0$ and $x^7=L$. This will result in multiplication of the action by a factor of 2 since the interval $I$ is half the size of the circle $S_1$ . Assuming that all fields are continuous and smooth, the parity assignments then imply the following boundary conditions:
\be
\bs
 (\hat{e}_{\mu}{}^{\usevenn},\hat{e}_{7}{}^{\unn}, \hat{A}_{\mu \nu \rho}, \hat{A}_\mu^I, A_7^{I^\prime}, \phi^{r'i}, \hat\psi_{\um -}, \hat\psi_{\usevenn +}, \hat\chi_+, \hat\lambda^r_+, \hat\lambda^{r^\prime}_-)\big |_{\partial_M} &= 0\ ,
\w2
\partial_7 ( \hat{e}_{\mu}{}^{\unn}, \hat{e}_{7}{}^{\usevenn} , \hat{A}_{ \mu \nu 7 }, \hat\sigma, \hat{A}_\mu^{I^\prime}, \hat{A}_7^I, \phi^{ri}, \hat\psi_{\um +}, \hat\psi_{\usevenn -}, \hat\chi_-, \hat\lambda^r_-, \hat\lambda^{r^\prime}_+){\big |}_{\partial M} &= 0\ .
\label{emptybc}
\es
\ee
The boundary conditions on $\phi$--scalars imply that the even-parity coset representatives $(L_I^i, L_I^r)$ parametrize the coset $SO(p,3)/SO(p)\times SO(3)$, and $L_{I'}^{r'}=\delta_{I'}^{r'}$, whilst the odd-parity coset representatives $(L_{I'}^i, L_{I'}^r, L_I^{r'})$ vanish on the boundaries.

The fields whose $\partial_7$ derivatives vanish at the boundaries are the parity even ones. In a diagonalised basis which will be spelled out in the next section (see eqn. \eq{redef}), they arrange themselves into $D=6$ supergravity plus a single tensor multiplet, $(n-p)$ vector multiplets and $(p+1)$ hypermultiplets.

We also note that our parity assignments differ from those used in \cite{Gherghetta:2002nq,Avramis:2004cn} in two respects. Firstly, while the coupling constant $g$ is declared to be parity odd in \cite{Gherghetta:2002nq,Avramis:2004cn}, we take it here to be parity even. Secondly, while all the vector fields are taken to be parity odd in \cite{Gherghetta:2002nq,Avramis:2004cn}, here we split them into two sets, and we assign even parity to one of these sets. Both of these differences crucially depend on our working with a noncompact gauged supergravity in $D=7$.

In order that the Euler-Lagrange variational principle be consistent with these boundary conditions, the action has to be supplemented by suitable additional terms defined on the boundary, known as Gibbons-Hawking-York terms. Then the total action takes the form
\begin{equation}
S = \int_M d^7x \mathcal{L}_{SG} + \int_{\partial M} d^6x \mathcal{L}_{GHY}\ .
\end{equation}
In the rest of this section, we will determine ${\cal L}_{GHY}$. We will consider explicitly the boundary at $x^7 = 0$. The boundary located at $x^7=L$ can be treated similarly.

To begin with, let us consider a general variation of the Einstein-Hilbert term. It contains a normal derivative of the metric variation, which must be avoided in order that the boundary conditions implied by the variational principle are not over constrained. To achieve this, as is well known, one adds an extrinsic curvature term so that the total action becomes \footnote{We could alternatively have defined $\hat{R}$ with respect to the spin connection which would then contain fermi squared terms. However that definition contributes a total derivative which is subsequently eliminated by adding appropriate Gibbons-Hawking-York terms, with no further effect in the bulk plus boundary theory that we will construct \cite{Moss:2004ck}.}
\begin{equation}
S_{EH} +S^0_{GHY} = \frac{1}{2\kappa^2} \int_M d^7x \hat{e} \hat{R} +
\frac{1}{\kappa^2} \int_{\partial M} d^6x \sqrt{-\hat{h}} \hat{K}\ ,
\label{EH}
\end{equation}
where $\hat{K}$ is the extrinsic curvature, which is defined as follows.
Let  $\hat{n}_N$ denote the unit vector normal to the boundary pointing out of $M$. We construct the induced metric $\hat{h}_{MN}$ as
\begin{equation}
 \hat{g}_{MN} = \hat{h}_{MN} + \hat{n}_{M} \hat{n}_{N}\ ; \qquad \hat{n}^M\hat{h}_{MN}=0\ .
\label{defh}
\end{equation}
Consequently, contraction with $\hat{h}_{MN}$ projects onto components of vectors in directions tangent to the boundary. The extrinsic curvature
is defined as
\begin{equation}
\hat{K} = \hat{h}^{MN} \hat{K}_{MN}\ , \quad \hat{K}_{MN}= \hat{h}_M^P \hat{h}_N^Q \hat{\nabla}_P \hat{n}_Q\ .
\end{equation}
Then the general variations of \eqref{EH} yields, modulo the Einstein field equation,
\be
\left(\delta S_{EH} + \delta S_{GHY}^0\right) |_{EOM} = - \frac1{2\kappa^2} \int_{\partial M} dx^6  \sqrt{-\hat{h}} \left( \hat{K}^{MN} - \hat{K}\hat{h}^{MN} \right) \delta \hat{g}_{MN}\ .
\label{varEHGHhat}
\ee
This vanishes, however, upon imposing the boundary conditions \eq{emptybc}, which in particular imply
\be
K_{\mu\nu}|_{\partial M}=0\ .
\label{k}
\ee
Turning to the general variation of the fermionic kinetic terms, they all involve fermion variations of both chiralities. In order that the boundary conditions implied by the variational principle  are not over constrained, we add suitable Gibbons-Hawking terms such that
\be
\bs
S_{F} +S^1_{GHY} &= \frac{1}{\kappa^2} \int_M  d^7x \hat{e} \bigg \lbrace -  \frac{i}{2} \hat{\bar{\psi}}_M \hat{\gamma}^{MNR} \hat{D}_N \hat{\psi}_R - \frac{5i}{2}  \hat{ \bar \chi} \hat{\gamma}^M  \hat{D}_M \hat \chi - \fr{i}{2g^2} \hat{\bar \l}^{\hat{r}} \hat{\g}^{M} \hat{D}_{M} \hat{\l}_{\hat{r}} \bigg \rbrace
\w2
& +\frac{1}{\kappa^2} \int_{\partial M} d^6 x \sqrt{-\hat{h}} \bigg \lbrace  -\frac{i}{4} \hat{\bar\psi}_\mu \hat{\gamma}^{\mu \nu} \hat\psi_\nu  - \frac{5i}{4} \hat{\bar\chi} \hat\chi - \fr{i}{4g^2} \hat{\bar \l}^{r} \hat{\l}_{r} + \fr{i}{4g^2} \hat{\bar \l}^{r\pr} \hat{\l}_{r\pr}\bigg \rbrace\ .
\label{defF}
\es
\ee
As a result, we obtain the total variation, modulo the fermion equations of motion,
\be
\bs
\left(\delta S_{F} + \delta S_{GHY}^1 \right)|_{EOM} & = \frac{1}{\kappa^2} \int_{\partial M} d^6 x \sqrt{-\hat{h}} \bigg \lbrace -  i \hat{\bar\psi}_{\mu - } \hat{\gamma}^{\mu \nu} \delta \hat\psi_{\nu + } \\
& - 5 i \hat{ \bar \chi}_+ \delta \hat \chi_-
- \fr{i}{g^2} \hat{\bar \l}^{r}_+ \d \hat{\l}_{r -} + \fr{i}{g^2} \hat{\bar \l}^{r\pr}_- \d \hat{\l}_{r\pr +}
\bigg \rbrace\ ,
\label{varRSGH}
\es
\ee
which is set to zero when the parity-odd fields vanish on the boundary.

One can check that there is no need for any further Gibbons-Hawking terms, and we conclude that the total action with a well-defined variational principle yielding the bulk equations of motion and the boundary conditions \eq{emptybc} is given by $S_{SG}+S^0_{GHY}+S^1_{GHY}$.

\section{Dimensional Reduction and the Diagonalised Basis for Fields}\label{7to6}

In describing the coupling of matter fields to supergravity on the boundary, which we shall do in the next section, it is convenient to express the parity-even bulk fields in a diagonal basis upon restriction to the boundary. In particular, the gravitino and dilaton field equations will be put into a canonical form in this basis. To achieve this, we shall consider the dimensional reduction of $S_{SG}$ on a circle and then will chirally truncate the theory such that we retain only the even-parity fields. This amounts to taking a limit in which the boundaries are empty and coincident, which results in a $D=6,\ N=(1,0)$ supergravity.

We begin by making a Kaluza-Klein ansatz for the the metric,
\begin{equation}
\hat{g}_{MN} = \left( \begin{array}{cc} e^{2 \alpha \phi} g_{\mu \nu} + e^{2 \beta \phi} \mathcal{A}_\mu \mathcal{A}_\nu & -e^{2 \beta \phi}  \mathcal{A}_\mu \\
-e^{2 \beta \phi}  \mathcal{A}_\mu & e^{2 \beta \phi} \end{array} \right)\ .
\label{defKaluza}
\end{equation}
We chose values for $\alpha$ and $\beta$ so as to obtain the standard Einstein-Hilbert gravitational action in $D=6$,
\begin{align}
\alpha &= - \frac{1}{2 \sqrt{10}}\ , & \beta &= - 4 \alpha\ .
\end{align}
We will chose our notation such that hatted fields have their indices raised and lowered with $\hat{g}_{MN}$, while unhatted fields have their indices raised and lowered with $g_{\mu \nu}$. We work with the corresponding vielbein basis,
\ba
\hat{e}^{\um}_\mu &= e^{ \alpha \phi} e^{\um}_\mu\ , &
\hat{e}^\mu_{\um} &= e^{- \alpha \phi} e^\mu_{\um}\ ,\nn \\
\hat{e}^\usevenn_\mu &= - e^{\beta \phi} \mathcal{A}_\mu\ , &
\hat{e}^7_{\um} &= e^{- \alpha \phi} \mathcal{A}_{\um}\ , \nn \\
\hat{e}^{\um}_7 &= 0\ ,&
\hat{e}^\mu_\usevenn &= 0 \ ,\nn \\
\hat{e}^\usevenn_{7} &= e^{\beta \phi}\ ,&
\hat{e}^7_\usevenn &= e^{- \beta \phi}\ .
\label{LorentzGauge}
\ea
We note here that in order for the gauge choice \eqref{LorentzGauge} to be invariant under the supersymmetry transformations \eqref{7DSUSY}, we must make a compensating Lorentz transformation with parameter $\lambda_{\um \useven} = - i \bar \epsilon_+  \g_{\um} \psi_{\useven +}$. As the veilbein is the only boson that transforms under Lorentz symmetry, the effect of this additional transformation on all other fields can be ignored, since it is higher order in fermions.

Working in a frame in which $\hat{n}_{\uM} = ( 0, 0, 0, 0, 0, 0, -1) $ implies that $\hat{n}_M = - \hat{e}^\usevenn_M$. Substituting this into \eqref{defh} we see that,
\begin{equation}
\hat{g}_{MN} = \left( \begin{array}{cc} \hat{h}_{\mu \nu} + e^{2 \beta \phi} \mathcal{A}_\mu \mathcal{A}_\nu & \hat{h}_{\mu 7} -e^{2 \beta \phi}  \mathcal{A}_\mu \\
\hat{h}_{\mu 7} -e^{2 \beta \phi}  \mathcal{A}_\mu & \hat{h}_{77} + e^{2 \beta \phi} \end{array} \right)\ .
\label{Kaluzah}
\end{equation}
Comparing \eqref{defKaluza} and \eqref{Kaluzah}, we can read off the components of $\hat{h}$ as
\begin{equation}
\hat{h}_{\mu \nu} = e^{2 \alpha \phi} g_{\mu \nu}, \quad \hat{h}_{\mu 7} = \hat{h}_{77} = 0\ ;
\end{equation}
this will be useful when determining the surface variations later on.

We can now diagonalise all kinetic terms by making the following redefinitions \cite{Bergshoeff:2005pq}
\begin{align}
\sigma &= \hat\sigma - 2 \alpha \phi\ , & \varphi &= \frac12 \hat\sigma + 4 \alpha \phi\ , \nonumber\\
\psi^r &= \frac{1}{g} \frac{1}{\sqrt{2}} e^{\frac{\alpha \phi}{2}} \hat\lambda^r_-\ , & \lambda^{r^\prime} &= \frac{1}{\sqrt{2}} e^{\frac{\alpha \phi}{2}} \hat\lambda^{r^\prime}_+\ , \nonumber \\
\chi &= \sqrt{2} e^{\frac{\alpha \phi}{2}} \left( \hat\chi_- + \frac14 \hat\psi_{\usevenn-} \right)\ , & \psi &= \frac{1}{\sqrt{2}} e^{\frac{\alpha \phi}{2}} \left( \hat\psi_{\usevenn-} - \hat \chi_- \right)\ , \nonumber\\
\psi_{\um} &= \frac{1}{\sqrt{2}} e^{\frac{\alpha \phi}{2}} \left( \hat\psi_{\um+} - \frac14 \gamma_{\um} \hat\psi_{\useven-} \right), & \hat\epsilon_+ &= \frac{1}{\sqrt{2}} e^{ \frac{\alpha \phi}{2}}  \epsilon\ , \label{redef}\\
\Phi^I &= \frac{1}{g} \hat{A}_7^I\ , & A_\mu^{I^\prime} &  = \hat{A}_\mu^{I^\prime}\ , \nonumber\\
B_{\mu \nu} &= \frac1{\sqrt2} \hat{A}_{\mu\nu 7}\ .\nonumber
\end{align}
With these definitions, the $D=6$ supergravity action becomes
\be
\bs
S_{SG (6)} =& \frac{2L}{{\kappa}^2} \int  d^6x e \bigg \lbrace \frac14 {R} - \frac1{8g^2} e^{{\sigma}}
{F}_{\mu \nu}^{r^\prime} {F}^{\mu \nu r^\prime } - \frac{1}{12} e^{-2 {\sigma}} {G}_{\mu\nu\rho} {G}^{\mu\nu\rho} - \frac14 \partial_\mu {\sigma} \partial^\mu {\sigma}\\
& - \frac14 \partial_\mu {\varphi} \partial^\mu {\varphi}  - \frac14 P_\mu^{i {r}} P^{\mu i {r}} - \frac14 {\cal P}_\mu^{{r}} {\cal P}^{\mu {r}} - \frac14 \mathcal{P}_\mu^{i } \mathcal{P}^{\mu {i}}\\
&- \frac18 g^2 e^{-\sigma} \left( C^{i {r^\prime}} C^{i {r^\prime}} + 2S^{i {r^\prime}} S^{i {r^\prime}} \right) + \frac{1}{24 g^2} \varepsilon^{\mu\nu\rho\sigma\lambda\tau} G_{\mu \nu \rho} \omega^0_{\sigma \lambda \tau} (A^{r^\prime}) \\
& -\frac{i}{2} {\bar{\psi}}_\mu {\gamma}^{\mu\nu\rho} {D}_\nu {\psi}_\rho - \frac{i}{2}  {\bar{ \chi}} {\gamma}^\mu {D}_\mu \chi - \frac{i}{2g^2} {\bar{\lambda}}^{{r^\prime}} {\gamma}^\mu {D}_\mu {\lambda}_{{r^\prime}} - \frac{i}{2}  {\bar{ \psi}} {\gamma}^\mu {D}_\mu \psi- \frac{i}{2}  {\bar{ \psi}}^{r} {\gamma}^\mu {D}_\mu \psi^{r}\\
& - \frac12 {\bar{\psi}}^{{r}} \sigma^i {\gamma}^\mu {\gamma}^\nu {\psi}_\mu P_\nu^{i {r}} - \frac12 {\bar{\psi}} \sigma^i {\gamma}^\mu {\gamma}^\nu {\psi}_\mu \mathcal{P}_\nu^{i}- \frac{i}{2}  {\bar{\psi}}^{{r}} {\gamma}^\mu {\gamma}^\nu {\psi}_\mu {\cal P}_\nu^{ {r}}\\
&- \frac{i}{2} {\bar{\chi}} {\gamma}^\mu {\gamma}^\nu {\psi}_\mu \partial_\nu {\sigma}  - \frac{i}{2} {\bar{\psi}} {\gamma}^\mu {\gamma}^\nu {\psi}_\mu \partial_\nu {\varphi} - \frac{i}{24 } e^{{-\sigma}} {G}_{\mu\nu\rho} \bigg (\bar\psi_{[\lambda} \gamma^\lambda \gamma^{\mu\nu\rho} \gamma^\tau \psi_{\tau]}\\
& -2 \bar \psi_\lambda \gamma^{\mu\nu\rho} \gamma^\lambda \chi - \bar\chi \gamma^{\mu\nu\rho} \chi + \bar\psi \gamma^{\mu \nu \rho} \psi + \bar\psi^r \gamma^{\mu\nu\rho} \psi^r - \frac1{g^2} \bar\lambda^{r^\prime} \gamma^{\mu\nu\rho} \lambda^{r^\prime} \bigg) \\
& - \frac14 \mathcal{P}_\mu^i \bigg( {\bar\psi}_{[\rho} \sigma^i  \gamma^{\rho} \gamma^{\mu} {\gamma}^\tau \psi_{\tau]} + {\bar\chi} \sigma^i \gamma^{\mu} {\chi} + \frac1{g^2} {\bar\lambda}^{{r}^\prime} \sigma^i \gamma^{\mu} \lambda^{{r}^\prime} -{\bar\psi}^{{r}} \sigma^i \gamma^{\mu} \psi^{{r}} - {\bar\psi} \sigma^i \gamma^{\mu} \psi \bigg) \\
&- i {\cal P}_\mu^r \bar\psi \gamma^\mu \psi^r - \frac{i}{4g^2} e^{\frac{{\sigma}}{2}} {F}_{\mu\nu}^{{r^\prime}} \bigg( {\bar\psi}_\rho \gamma^{\mu\nu} \gamma^{\rho} \lambda^{{r^\prime}} + {\bar\chi} \gamma^{\mu\nu} \lambda^{{r}^\prime} \bigg) \\
& - e^{\frac{\sigma}{2}} C^{i r r^\prime} \bar\lambda^{r^\prime} \sigma^i \psi^r + i e^{\frac{\sigma}{2}} S^{r r^\prime} \bar\lambda^{r^\prime} \psi^r - e^{\frac{\sigma}{2}} S^{i r^\prime} \bar\lambda^{r^\prime} \sigma^i \psi\\
&+ \frac{1}{2 \sqrt{2}} e^{-\frac{{\sigma}}{2}} \lambda^{r^\prime} \sigma^i \gamma^\mu \psi_\mu \left( C^{i {r^\prime}} - \sqrt{2} S^{i r^\prime} \right) + \frac{1}{2 \sqrt{2}} e^{-\frac{{\sigma}}{2}} \lambda^{r^\prime} \sigma^i \chi \left( C^{i {r^\prime}} - \sqrt{2} S^{i r^\prime} \right)\bigg \rbrace\ ,
\label{6DAction}
\es
\ee
where $\varepsilon^{\um\un\urho\usigma\ulambda\utau} = \hat\varepsilon^{\um\un\urho \usigma\ulambda\utau \useven}$\ , and we have used the following definitions:
\bea
{G}_{\mu\nu\rho} &=& 3 \partial_{[\mu} {B}_{\nu\rho]}\ ,\quad\quad
{F}^{{r^\prime}}_{\mu\nu} = 2 \partial_{[\mu} {A}_{\nu]}^{{r^\prime}} + {f_{{s^\prime} {t^\prime}}}^{{r^\prime}} {A}_\mu^{{s^\prime}} {A}_\nu^{{t^\prime}}\ ,
\nn\w2
\omega^0_{\mu\nu\rho} (A^{r^\prime}) &=& F_{[\mu\nu}^{r^\prime} A^{r^\prime}_{\rho]} - \frac13 f_{r^\prime s^\prime t^\prime} A^{r^\prime}_\mu A^{s^\prime}_\nu A^{t^\prime}_\rho\ ;
\eea
the elements of the Maurer-Cartan forms are defined as
\bea
P_\mu^{i {r}} &=& L^{{I} {r}} \left( \delta_{{I}}^{{K}} \partial_\mu -  {f_{{r^\prime} {I}}}^{{K}} {A}_\mu^{{r^\prime}} \right) L_{{K}}^i\ ,
\nn\w2
Q_\mu^{i j} &=& L^{{I} j} \left( \delta_{{I}}^{{K}} \partial_\mu -  {f_{{r^\prime} {I}}}^{{K}} {A}_\mu^{{r^\prime}}\right) L_{{K}}^i\ ,
\nn\w2
Q_\mu^{{r} {s}} &=& L^{{I} {r}} \left( \delta_{{I}}^{{K}} \partial_\mu -  {f_{{r^\prime} {I}}}^{{K}} {A}_\mu^{{r^\prime}} \right) L_{{K}}^{{s}}\ ,
\label{PQQ}
\eea
the axion field strengths are defined as
\bea
\mathcal{P}_\mu^i &=& e^\varphi \left( \pa_\mu \Phi^I + f^I{}_{r^\prime J} A_\mu^{r^\prime} \Phi^J \right) L_I^i\ ,
\nn\w2
\mathcal{P}_\mu^r &=& e^\varphi \left( \partial_\mu \Phi^I + f^I{}_{r^\prime J} A_\mu^{r^\prime} \Phi^J \right) L_I^r\ ,
\eea
gauge functions are defined as
\ba
C^{k r^\prime}  &= \frac{1}{\sqrt{2}} \epsilon^{kij} f_{r^\prime IJ} L^{Ii} L^{Jj}\ ,&
C^{irr^\prime}  &= f_{r^\prime IJ} L^{Ii} L^{Jj}\ \ ,
\nn\w2
S^{ir^\prime} &= - e^{\varphi} f_{r^\prime IJ} \Phi^J L^{Ii}\ , &
S^{r r^\prime} &= - e^\varphi f_{r^\prime IJ} \Phi^J L^{Ir}\ ,
\ea
and the covariant derivative is defined as
\be
{D}_\mu\epsilon = \left({\partial}_\mu  + \frac{1}{4} \omega_{\mu \um \un } \gamma^{\um\un } + \frac{1}{2 \sqrt{2}} Q_\mu^i \sigma^i\right)\epsilon\ , \quad\quad Q_\mu^i = \frac{i}{\sqrt{2}} \epsilon^{ijk} Q_\mu^{jk}\ .
\label{covder}
\ee
Truncating the supersymmetry transformations \eqref{7DSUSY} and writing the result in terms of the redefined fields gives the transformations under which the action \eqref{6DAction} is invariant:
\begin{equation}
\begin{split}
\delta {e}_\mu^{\um} &= i {\bar\epsilon} \gamma^{\um} \psi_\mu\ ,\\
\delta {\psi}_\mu &=  {D}_\mu \epsilon - \frac{i}{2} \mathcal{P}_\mu^i \sigma^i \epsilon\ + \frac{1}{24} e^{-\sigma} G_{\rho\sigma\tau}  \gamma^{\rho\sigma\tau} \gamma_\mu \epsilon\ , \\
\delta \chi &= - \frac12 \gamma^\mu \partial_\mu \sigma  \epsilon - \frac{1}{12} e^{-\sigma} {G}_{\mu\nu\rho} \gamma^{\mu\nu\rho} \epsilon\ ,\\
\delta {B}_{\mu\nu} &= - i e^{\sigma} \bar\epsilon \gamma_{[\mu} \psi_{\nu]} + \frac{i}2 e^\sigma \bar\epsilon \gamma_{\mu\nu} \chi\ ,\\
\delta \sigma &= - i {\bar\epsilon} \chi\ ,\\
\delta {A}_\mu^{{r^\prime}} &=  i e^{-\frac{\sigma}{2}} {\bar{\epsilon}} \gamma_\mu \lambda^{{r^\prime}}\ ,\\
\delta \lambda^{r^\prime} &= -\frac14 e^{\frac{\sigma}{2}} \gamma^{\mu \nu} F^{r^\prime}_{\mu \nu} \epsilon - \frac{i}{2 \sqrt{2}} g^2  e^{-\frac{\sigma}{2}} \left( C^{i r^\prime} - \sqrt{2} S^{i r^\prime} \right) \sigma^i \epsilon\ ,\\
\delta \psi &= \frac{i}{2} \gamma^\mu \left( \mathcal{P}_\mu^i \sigma^i - i \partial_\mu \varphi \right) \epsilon\ ,\\
\delta \psi^r &= \frac{i}{2} \gamma^\mu \left( P_\mu^{i r} \sigma^i + i \mathcal{P}_\mu^r \right) \epsilon\ , \\
\delta \varphi &= i \bar{\epsilon} \psi\ ,\\
\delta L_I^r &= \bar{\epsilon} \sigma^i \psi^r L_I^i\ ,\\
\delta L_I^i &= \bar{\epsilon} \sigma^i \psi^r L_I^r\ ,\\
\delta \Phi^I &= - L^{Ii} e^{-\varphi} \bar{\epsilon} \sigma^i \psi - i L^{Ir} e^{-\varphi} \bar{\epsilon} \psi^r\ .\\
\end{split}
\label{6DSUSY}
\end{equation}
The fields appearing here can be written in terms of $N=(1,0)$ multiplets in $D=6$. These consist of the supergravity multiplet $(e_\mu^{\um}, \psi_\mu, B_{\mu \nu}^+)$, a single tensor multiplet $( B_{\m \n}^-, \c, \s)$, vector multiplets $(A_\mu^{r\pr}, \l^{r\pr})$ and hypermultiplets $(L_I^r, L_I^i, \F^I, \vf, \p, \p^r)$.  By making suitable redefinitions, it is possible to demonstrate that the scalars of the hypermultiplets form the enlarged coset
\be
\fr{SO(p+1,4)}{SO(p+1)\times SO(4)}
\ee
which is a quaternionic K\"{a}hler manifold \cite{Bergshoeff:2005pq}. However we will not make these redefinitions here.

These redefined fields and transformations represent the induced supergravity which is present on the boundary and it is to this supergravity that we will couple boundary-localised matter in the following sections. When the boundaries are populated by this localised matter, the transformations \eqref{6DSUSY} will be modified corresponding to non-zero odd $\times$ odd terms appearing in the variation of these even-parity fields. However these transformations will be of higher order in the boundary couplings and so will be ignored in this paper.

\section{Introduction of Boundary Yang-Mills Fields and the Modified Boundary Conditions}\label{BCs}

We will now consider turning on a boundary action describing vector multiplets
\begin{equation}
( C_\mu^X , \eta^{XA} )\ ,
\end{equation}
\smallskip
where $\eta$ is an $Sp(1)$ pseudo-Majorana spinor with a doublet index $A$ as before and $X$ labels the adjoint representation of some gauge group $K'$.
The supersymmetry transformations of these boundary fields must be given by their known flat-space forms modified by appropriate bulk dressings. We therefore make the ansatz,
\begin{equation}
\begin{split}
\delta C_\mu^X &= i e^{-  \frac{a\sigma}{2}}  \bar \epsilon \gamma_\mu \eta^X\ ,\\
\delta \eta^X &= - \frac14 e^{\frac{a\sigma}{2}}  \gamma^{\mu \nu} H_{\mu \nu}^X \epsilon\ ,
\end{split}
\label{VecSUSY}
\end{equation}
where $H_{\mu \nu}^X=\partial_\mu C_\nu^X-\partial_\nu C_\mu^X + f^{X}{}_{YZ} C_\mu^YC_\nu^Z$ and $a$ is a constant which is to be determined.  From our analysis in Section \ref{7to6}, we recognise that the scalar $\vf$ forms part of the $D=6$ quaternionic K\"{a}hler coset, and as such it is does not arise in the above transformation rules.

An immediate consequence of having introduced a boundary action is the modification of the boundary condition \eq{k} such that $K_{\mu\nu}-g_{\mu\nu}K$ will now be proportional to the stress tensor of the boundary action. This condition is known as the Israel junction condition \cite{Israel:1966rt}. On the other hand, since the supersymmetry transformation of the odd-parity gravitino $\psi_{\mu -}$ contains the extrinsic curvature $K_{\mu\nu}$, it follows that we must modify its boundary condition too. Supersymmetry will then require that we modify other boundary conditions as well. To determine these modifications, we begin by recording the supersymmetry transformation rules of the parity-odd fields\footnote{For clarity, these have been truncated to include only parity-odd fields that receive nontrivial boundary conditions in the following analysis.}
\begin{equation}
\begin{split}
\delta \psi_{\mu-} &= - \frac{1}{2} K_{\mu \nu} \gamma^{\nu} \epsilon - \frac{1}{480} e^{ - \sigma } F_{\rho \sigma \lambda \tau} (\gamma_{\mu} \gamma^{\rho \sigma \lambda \tau} + 5\, \gamma^{\rho \sigma \lambda \tau} \gamma_{\mu}) \epsilon\ , \\
\delta \chi_+ &=  - \frac{1}{4} e^{ 6 \alpha \phi} \partial_{ \bar{ 7} } \hat{\sigma}  \epsilon - \frac{1}{120} e^{ - \sigma}  F_{ \mu \nu \rho \sigma} \gamma^{\mu \nu \rho \sigma} \epsilon \ ,
\end{split}
\label{7DSUSYodd}
\end{equation}
where we have used the bulk supersymmetry transformations \eqref{7DSUSY} and have made the following redefinitions
\begin{align}
K_{\mu \nu} &= e^{4 \alpha \phi} \hat{K}_{\mu \nu}\ , & K &= K_{\mu \nu} g^{\mu \nu}\ , \nonumber\\
\psi_{\mu -} &= \frac1{\sqrt{2}} e^{\frac{9 \alpha \phi}{2}} \hat\psi_{\mu -}\ , & \chi_+ &= \frac{1}{\sqrt{2}} e^\frac{11 \alpha \phi}{2} \hat\chi_+\ , \label{redefodd}\\
F_{\mu\nu\rho\sigma} &= \frac1{\sqrt{2}} \hat{F}_{\mu\nu\rho\sigma}\ .  \nonumber
\end{align}
We have also used the identity
\begin{equation}
P_- \left( \hat D_\mu \hat\epsilon \right) = - \hat D_\mu \left( P_- \right) \hat \epsilon =  -\frac1{2 \sqrt{2}} K_{\mu \nu} \gamma^\nu e^{- \frac{9 \alpha \phi}{2} } \epsilon\ .
\end{equation}
Examining these transformations, it follows that we need also to specify the modified boundary conditions for $F_{\mu\nu\rho\sigma}$, $\partial_{\usevenn}{\hat\sigma}$ and $\chi_+$ in a manner consistent with \eq{7DSUSYodd}. Carrying out this process yields the modified boundary conditions
\smallskip
\bea
\psi_{\mu-}\ev &=& - \frac{7}{20} b e^{ (c + \frac{a}{2} ) \sigma} H_{\mu \nu}^X \gamma^\nu \eta_X + \frac{3}{40} b e^{(c + \frac{a}{2} ) \sigma} H^{\rho \sigma X}  \gamma_{\mu \rho \sigma} \eta_X +  (\text{fermi})^3\ ,
\nn\w2
\chi_+\ev &=& \frac1{20} b  e^{(c + \frac{a}{2} ) \sigma} H_{\mu \nu}^X \gamma^{\mu \nu} \eta_X  +  (\text{fermi})^3\ ,
\nn\w2
e^{6 \alpha \phi} \partial_{\usevenn} \hat{\sigma}\ev &=&  - \frac{1}{10} b e^{( c+a) \sigma} H_{\mu \nu}^X H^{\mu \nu}_X  +  (\text{fermi})^2\ ,
\label{VectorBC}\w2
F_{\mu \nu \rho \sigma}\ev &=&  \frac32 b e^{ (1+ c+ a)\sigma}  H_{[\mu\nu}^X H_{\rho\sigma] X}  + (\text{fermi})^2\ ,
\nn\w2
K_{\mu\nu}\ev &=& \frac12 b e^{ (c + a) \sigma} H_{\mu \rho}^X H\indices{_\nu^\rho_X} - \frac3{40}  b e^{( c+ a )\sigma} H_{\rho\sigma}^X H^{\rho\sigma}_X g_{\mu\nu}  + (\text{fermi})^2 \ ,
\nn
\eea
\smallskip
where $b$ and $c$ are further constants, which will be determined in the next section by considering the cancellation of certain terms in the supersymmetry variation. Furthermore, the bulk Bianchi identity $\partial_{[\mu} \hat{F}_{\nu \rho \sigma \tau]}= 0$ implies that  $1+a +c=0$. The boundary conditions on all other parity-odd bulk fields vanish at lowest order in fermions.

We can rephrase the boundary condition on $F_{\mu\nu\rho\sigma}$ in terms of a condition on $A_{\mu\nu\rho}$.  However, in order to do this we must first modify the bulk supersymmetry transformation of $\hat A_{MNR}$ to
\begin{equation}
\delta \hat A_{MNR} = \frac{3i}{\sqrt 2} e^{\hat \sigma} \hat {\bar\epsilon} \hat \gamma_{[M N} \hat \psi_{R]} - i \sqrt{2}  e^{\hat \sigma} \hat {\bar\epsilon} \hat \gamma_{MNR} \hat \chi + \partial_{[ M} \hat f^1_{NR]}\ .
\label{7DSUSYA}
\end{equation}
Here, $\hat f^1_{NR}$ is an arbitrary function, linear in $\hat \epsilon$. This does not effect the bulk supersymmetry as $ \hat A_{MNR}$ always appears through $\hat F_{MNRS}$ or multiplies a total derivative in \eqref{7DAction}.
Making an ansatz for the boundary condition on $A_{\mu \nu \rho}$ and then enforcing that its variations under \eqref{VecSUSY} and \eqref{7DSUSYA} match, we find that
\ba
A_{\mu\nu\rho}\ev &\equiv \frac{1}{\sqrt{2}} \hat{A}_{\mu\nu\rho} \ev \nn\\
&=  \frac{3}{4} b \omega^0_{ \mu\nu\rho} (C)  + \frac{i}{8} b e^{- a \sigma} {\bar \eta^X} \gamma_{\mu\nu\rho} \eta_X \ ,
\label{ABC2}
\ea
and
\be
f^1_{\mu \nu}\ev \equiv \frac1{\sqrt{2}} \hat f^1_{\mu \nu}\ev = \frac32 b \delta_\epsilon C_\mu^X C_{\nu X}\ .
\label{f1}
\ee
Consistency with the boundary Yang-Mills gauge transformations then requires that we impose the following boundary condition on the tensor gauge transformation parameter
\begin{equation}
\lambda_{\mu \nu}\ev \equiv \frac1{\sqrt{2}}\hat\lambda_{\mu \nu}\ev=\frac{1}{2} b \partial_{[\mu} C_{\nu]}^X \Lambda_X\ .
\label{LambdaBC}
\end{equation}
As we shall see later, the boundary conditions \eq{f1} and \eq{LambdaBC} will play a crucial role in the identifications of the supersymmetry and gauge anomalies, respectively. Note also that in determining \eq{f1}, we needed to include the term bilinear in fermions. While we did not need to specify the bilinear fermion terms in \eq{VectorBC} to the order to which we are working in determining the boundary action, there is a need to do so in the case of $A_{\mu}^{I}$ in studying the coincident-boundary limit of the bulk-plus-boundary system, as we shall see in Appendix \ref{coincident}. In that case, the appropriate boundary condition can be seen to be
\begin{equation}
A_\mu^I\ev = - \fr{\k^2}{4\l^2} e^{-\vf} \bar\h^X \s^i \g^\m \h_X L^{iI} + (\text{fermi})^4\ .
\end{equation}
Next, we shall construct the boundary Yang-Mills action, and we shall see that certain cancellations between the boundary action and the surface terms will fix the coefficients $a,b,c$, which are already subject to the condition $a+c+1=0$, as we have seen above.

\section{The Boundary Yang-Mills Action and Classical Anomalies}\label{BoundaryAction}

The general variation of the bulk action supplemented by the Gibbons-Hawking-York terms defined in \eq{EH} and \eq{defF} is given by
\ba
\delta S_{SG} &+ \delta S_{GHY}^0 + \delta S_{GHY}^1 = \int_M d^7x\, \hat{e}\delta \mathcal{L}_{(7)}
\label{varGHSG}\\
& + \frac{1}{\kappa^2} \int_{\partial M} d^6 x  \sqrt{-\hat{h}} \bigg \lbrace  - \frac12 \left( \hat{K}^{MN} - \hat{K} \hat{h}^{MN} \right) \delta \hat{g}_{MN}
\nn\\
&- i \hat{\bar\psi}_{\mu -} \hat\gamma^{\mu \nu}   \delta\hat\psi_{\nu+} - 5i \hat{ \bar\chi}_+  \delta \hat{\chi}_-  + \frac{5}{4} \partial_{\usevenn} \hat\sigma \delta \hat\sigma - \frac{1}{g^2}  e^{\hat{\sigma} } \delta \hat{A}_\mu^{I} L_{I}^{i} \hat{F}^{\mu \usevenn i}
\nn\\
& - \frac16 e^{- 2 \hat{\sigma}} \hat{F}^{\mu \nu \rho \usevenn} \delta \hat A_{\mu \nu \rho}
- \frac{1}{6 \sqrt{2}g^2} \hat{\varepsilon}^{\mu \nu \rho \sigma \lambda \tau} \hat{A}_{\mu \nu \rho} \hat{F}_{\sigma \lambda}^{r^\prime} \delta A_{\tau r^\prime} \bigg \rbrace\ ,
\nn
\ea
where all parity-odd fields other than those occurring in the modified boundary conditions \eqref{VectorBC} have been set to zero. It is important to note that we have performed an integration by parts in such a way that $\delta \mathcal{L}_{(7)}$ contains no derivatives of the variations.  However, in considering the variation of the bulk action under supersymmetry, which we shall do next, there will be extra surface terms due to the fact that further integrations by parts will be needed in order to leave the supersymmetry parameter undifferentiated. These are due to derivatives of $\epsilon$ present in the variation of the gravitino and the 3-from. Collecting the resulting surface terms, we find
\ba
\int_M  d^7x \delta_\epsilon \mathcal{L}_{(7)} &= \frac{1}{\kappa^2}\int_{\partial M} d^6 x \sqrt{ - \hat{h}} \hat{n}_M \bigg \lbrace
- 2i \hat{\bar\epsilon} \hat\gamma^{MNR} \hat{D}_N \hat\psi_R - \frac{i5}{2} \hat{\bar\chi} \gamma^M \gamma^N \hat\epsilon \hat \partial _N \hat \sigma
\nn\\
& + \frac{i}{96 \sqrt{2}} e^{-\hat{\sigma}} \hat{F}^{RSTU} \bigg ( 4 \hat{\bar\epsilon} \hat\gamma^{[M} \hat\gamma_{RSTU} \hat\gamma^{N]} \hat\psi_{N} + 8 \hat{\bar\epsilon} \hat{\gamma}_{RSTU} \hat\gamma^M \hat\chi \bigg )
\nn\\
& + \frac1{8g} e^{\frac{\hat\sigma}{2}} \hat{F}^{RSi} \left( 4 \hat{\bar\epsilon} \sigma^i \hat\gamma^{[M} \hat\gamma_{RS} \hat\gamma^{T]} \hat\psi_T - 4 \hat{\bar\epsilon} \sigma^i \hat\gamma_{RS} \hat\gamma^M \hat\chi \right)
\\
& +\frac16 e^{-2 \hat \sigma} \partial_N \hat f^1_{RS} \hat F^{MNRS} - \frac1{24 \sqrt{2}g^2 } \hat \varepsilon^{RSTUVWM} \hat f^1_{RS} \hat F_{TU}^{r^\prime} F_{VW}^{r^\prime}  \bigg\rbrace\ .
\nn
\ea
Substituting this into \eqref{varGHSG} and imposing the boundary conditions gives, after some algebra,
{\ba
\delta_\epsilon S_{SG} &+ \delta_\epsilon S_{GHY} = \frac{1}{\kappa^2} \int_{\partial M} d^6x e b \bigg \lbrace - \frac{1}{8} e^{ - (1+\frac{a}{2}) \sigma} {\bar{\epsilon}} \gamma^{\rho \sigma} \gamma^{\mu} \sigma^i \eta^X H_{\rho \sigma X} \mathcal{P}_{\mu}^{i}
\label{SurfVar}\\
&+  \frac{i}{48} e^{ -( 2+ \frac{a}{2} ) \sigma }  \bar\epsilon \gamma^{\rho \sigma \tau} \gamma^{\mu \nu} \eta^X  H_{\mu\nu X} G_{\rho \sigma \tau}
-  \frac{i}{96} e^{-( 2+ \frac{a}{2} ) \sigma }  \bar\epsilon \gamma^\lambda \gamma^{\rho \sigma \tau} \gamma^{\mu \nu} \gamma_\lambda \eta^X  H_{\mu\nu X} G_{\rho \sigma \tau}
\nn\\
&+ \frac{i}{16} e^{ -\sigma} \bar\epsilon \gamma^{\mu \nu \rho \sigma \tau} \psi_\tau  H^X_{\mu \nu} H_{\rho \sigma X} + \frac{i}{16} e^{ - \sigma} {\bar \epsilon} \gamma^{\mu \nu \rho \sigma } \chi H_{\mu \nu}^X H_{\rho \sigma X}
\nn\\
&- \frac{1}{8g^2}\varepsilon^{\mu\nu\rho\sigma \lambda \tau} \omega_{\mu\nu\rho}^0 (C) F_{\sigma \lambda}^{r^\prime} \delta_\epsilon A_{\tau r^\prime} + \frac1{16g^2} \varepsilon^{\mu \nu \rho \sigma \lambda \tau} \delta_{\epsilon} C_\mu^X C_{\nu X}  F^{r^\prime}_{\rho \sigma} F^{r^\prime}_{\lambda \tau}  \bigg \rbrace\ ,
\nn
\ea}
where $S_{GHY}=S_{GHY}^0+S_{GHY}^1$ as defined in \eqref{EH} and \eqref{defF}.
Next we construct the boundary action such that, together with the bulk action and subject to the modified boundary conditions \eq{VectorBC}, the total action is invariant under supersymmetry except for the last two terms in \eq{SurfVar}, which will be interpreted as supersymmetry anomalies and will be discussed in more detail below.

After some algebra we find that the boundary action is given by
\begin{equation}
\begin{split}
S_{YM} = \frac1{\lambda^2} \int_{\partial M} d^6 x e \bigg \lbrace & - \frac18 e^{- \sigma} H_{\mu\nu}^X H^{\mu\nu}_X  - \frac{i}{2} \bar \eta^X \gamma^\mu D_\mu \eta_X \\
&- \frac{i}{4} e^{-\frac\sigma2} H_{\rho\sigma}^X \bar\eta_X \gamma^\mu \gamma^{\rho\sigma} \psi_\mu - \frac{i}{4} e^{- \frac\sigma2} H_{\mu\nu}^X \bar\eta_X \gamma^{\mu\nu} \chi \bigg \rbrace\ ,
\label{NoetherCoupledVectorAct}
\end{split}
\end{equation}
where we have determined that 
\be
a=-1 \quad \quad \text{and} \quad \quad b = \frac{\kappa^2}{\lambda^2}\ ,
\label{abfixed}
\ee
as required to ensure certain cancellations between the variations of the boundary action and the surface term.

One might have expected a term of the form $G_{\mu \nu \rho} \bar\h^X \g^{\m \n \r} \h_X$ to appear in the boundary action, as such a term is present in the $D=6$ actions of \cite{Nishino:1986dc, Nishino:1997ff} and was claimed to be present in \cite{Gherghetta:2002nq}. However the Noether procedure does not require such a term and thus it is absent in the boundary action that we have derived. In Appendix A we will demonstrate that this term emerges in the coincident-boundaries limit by considering the boundary condition $A_{\mu \nu \rho} \ev \sim \bar h^X \g_{\m \n\r} \h_X$.  In this limit the 4-form kinetic term $F_{MNRS} F^{MNRS}$ will then give rise to the required  term in the reduced action \eqref{6DAction2}. A similar process is also described in \cite{Moss:2004ck}. 

With the parameters $a,b$ fixed as in \eqref{abfixed} the completely determined boundary conditions take the form:
\begin{equation}
\begin{split}
\psi_{\mu-}\ev &=  - \frac{7\kappa^2}{20\lambda^2} e^{ -\frac{\sigma}{2} } H_{\mu \nu}^X \gamma^\nu \eta_X + \frac{3\kappa^2}{40\lambda^2} e^{-\frac{\sigma}{2}} H^{\rho \sigma X}  \gamma_{\mu \rho \sigma} \eta_X + (\text{fermi})^3\ ,
\\
\chi_+ \ev &= \frac{\kappa^2}{20\lambda^2} e^{-\frac{\sigma}{2}} H_{\mu \nu}^X \gamma^{\mu \nu} \eta_X + (\text{fermi})^3\ ,
\\
 e^{6 \alpha \phi} \partial_{\usevenn} \hat{\sigma}\ev &=  - \frac{\kappa^2}{10\lambda^2}  e^{-\sigma} H_{\mu \nu}^X H^{\mu \nu }_X + (\text{fermi})^2\ ,
 \\
A_{\mu\nu\rho}\ev &=  \frac{3\kappa^2}{4\lambda^2} \omega^0_{ \mu\nu\rho} (C)  + \frac{i\kappa^2}{8\lambda^2} e^{ \sigma} \bar\eta^X \gamma_{\mu\nu\rho} \eta_X + (\text{fermi})^4\ ,
\\
A_\mu^I \ev &= - \frac{\kappa^2}{4\lambda^2} e^{-\vf} \bar\h^X \s^i \g^\m \h_X L^{iI} + (\text{fermi})^4\ ,
\\
K_{\mu\nu}\ev &= \frac{\kappa^2}{2\lambda^2} e^{ - \sigma} H_{\mu \rho}^X H\indices{_\nu^\rho_X} - \frac{3\kappa^2}{40\lambda^2} e^{-\sigma} H_{\rho\sigma}^X H^{\rho\sigma }_X g_{\mu\nu} + (\text{fermi})^2\ .
\label{detBC}
\end{split}
\end{equation}
The boundary conditions on all other parity-odd fields in \eqref{odd}are set to zero at lowest order in fermions. The vanishing boundary conditions on $L_{I\pr}^i$, $L_{I\pr}^r$ and $L_I^{r\pr}$ imply that the parity-odd C-functions $C$ , $C^{i r}$, $C^{i r s}$ and $C^{i r\pr s\pr}$ are also set to zero on the boundary. We also note that in Ref.\ \cite{Gherghetta:2002nq}, only the boundary condition on $A_{\mu \nu \rho}$ was considered, while our boundary conditions correspond to the completion of this to a full orbit.

At this point, it is important to check that these boundary conditions are also consistent with the variational principle following from the bulk + boundary action $S=S_{SG}+S_{GHY}+S_{YM}$. For example, the variation of the gravitino gives the boundary contribution
\begin{equation}
\int_{\partial M} d^6 x  e \bigg \lbrace - \frac{2i}{\kappa^2}  \bar\psi_{\mu -} \gamma^{\mu \nu}  - \frac{i}{ 4 \lambda^2}  e^{-\frac{\sigma}{2}} H_{\rho\sigma}^X \bar{\eta}^X \gamma^\nu \gamma^{\rho \sigma}  \bigg \rbrace \delta\psi_{\nu}\ ,
\end{equation}
which is set to zero by imposing the boundary condition on $\psi_{\mu -} $ given above. Similarly, we have checked that the surface terms that arise in the variations of all the other fields cancel upon use of the stated boundary conditions and boundary field equations.

Next, we turn to the nonvanishing last two terms in \eq{SurfVar}, which we now identify as the residual supersymmetry anomaly. We note that there is also an anomaly in the boundary Yang-Mills transformation, and, together with the supersymmetry anomalies, they must together satisfy the Wess-Zumino consistency conditions. To see this in more detail, it is convenient to add the local counterterm
\begin{equation}
S_{YM}^{\prime} = \frac1{32\lambda^2 g^2} \int_{\partial M} d^6 x e \varepsilon^{\mu\nu\rho\sigma\lambda \tau}  \omega_{\mu\nu\rho}^0 (C) \omega_{\sigma\lambda\tau}^0 (A)\ .
\label{prime}
\end{equation}
This also produces a gauge anomaly in the bulk Yang-Mills gauge transformations and puts the total gauge anomaly into a symmetric form known as the consistent anomaly \cite{RicSag}. Then the total variation of the action $S'=S_{SG}+S_{GHY}+S'_{YM}$ under the Yang-Mills gauge transformations is given by
\begin{equation}
\delta_\Lambda S' = \frac1{32\lambda^2 g^2} \int_{\partial M} d^6 x e \bigg \lbrace  \varepsilon^{\mu \nu\rho \sigma \lambda \tau} H_{ \mu \nu}^X H_{\rho \sigma X} \partial_\lambda A_\tau^{r^\prime} \Lambda^{r^\prime} + \varepsilon^{\mu \nu\rho \sigma \lambda \tau} F_{ \mu \nu}^{r^\prime} F_{\rho \sigma}^{r^\prime} \partial_\lambda C_\tau^{X} \Lambda_{X} \bigg \rbrace
\label{AnomClasGaugeone}
\end{equation}
and the last two terms in \eq{SurfVar} together with the supersymmetry variation of \eq{prime} yield the corresponding supersymmetry anomaly
\begin{equation}
\begin{split}
\delta_{\epsilon} S' &= \frac{1}{32\lambda^2 g^2} \int_{\partial M} d^6 x e \bigg \lbrace \varepsilon^{\mu\nu\rho\sigma \lambda \tau} H_{\mu \nu}^X H_{\rho\sigma X} \delta_\epsilon A_\lambda^{r^\prime} A_\tau^{r^\prime}
- 2 \varepsilon^{\mu\nu\rho\sigma \lambda \tau} \omega_{\mu\nu\rho}^0 (C) F_{\sigma \lambda}^{r^\prime} \delta_\epsilon A_\tau^{r^\prime} \\
& + \varepsilon^{\mu\nu\rho\sigma \lambda \tau} F_{\mu \nu}^{r^\prime} F_{\rho\sigma}^{r^\prime} \delta_\epsilon C_\lambda^X C_{\tau X}
- 2 \varepsilon^{\mu\nu\rho\sigma \lambda \tau} \omega_{\mu\nu\rho}^0 (A) H_{\sigma \lambda}^X \delta_\epsilon C_{\tau X} \bigg \rbrace\ .
\end{split}
\label{AnomClasSUSYone}
\end{equation}
Finally, one may verify that these two anomalies indeed do satisfy the complete set of Wess-Zumino consistency conditions
\begin{align}
\delta_{\Lambda_1} \delta_{\Lambda_2} S' - \delta_{\Lambda_2} \delta_{\Lambda_1} S' &= \delta_{[\Lambda_1, \Lambda_2 ]} S'\ ,\\
\delta_\epsilon \delta_\Lambda S' - \delta_\Lambda \delta_\epsilon S' &= 0\ ,\\
\delta_{\epsilon_1} \delta_{\epsilon_2} S'- \delta_{\epsilon_2} \delta_{\epsilon_1}S' &= \delta_{\tilde{\Lambda}} S'\ ,
\end{align}
where $\tilde \Lambda$ is the gauge transformation produced by the commutator of two supersymmetry transformations in the standard way.

\section{Coupling Boundary Localised Hypermultiplets}\label{boundaryhypers}

Next, let us consider the coupling of boundary-localised hypermultiplets. We will carry out this coupling assuming no boundary-localised vector multiplets are present. These could be reintroduced later in order to gauge the hypermultiplet symmetries. The calculation will be similar to that carried out for vector multiplets in the previous sections. First we will find a supersymmetric set of boundary conditions, then we will construct the surface term produced upon varying the bulk action, and finally we will construct a boundary-localised action which varies to cancel this surface term.

We begin by considering $m$ hypermultiplets consisting of $4m$ real scalar fields $\phi^\alpha$ and symplectic Majorana-Weyl spinors $\zeta^a\,(a=1,...,2m)$. By global supersymmetry, it is known that the scalars must parametrize a hyperk\"ahler manifold ${\cal M}$, which is characterised by having a holonomy group $H$  contained in $Sp(m)$.  The scalar target manifold ${\cal M}$ may or may not have isometries. This will not play a role in our construction below. Let us denote the vielbeins on ${\cal M}$ by $V_\a^{a A}$. By supersymmetry, they must be covariantly constant
\be
\partial_\a V_{\b aA} - \Gamma_{\alpha\beta}^\gamma V_{\gamma aA} + \omega_{\a a}{}^b  V_{\b bA} +\omega_{\a A}{}^B V_{\b aB} =0\ ,
\ee
where $\Gamma_{\alpha\beta}^\gamma$ is the Levi-Civita connection, $\omega_\alpha^{ab}$ is an $H\subseteq Sp(m)$ valued connection and $\omega_\alpha^{AB}$ is an $Sp(1)_R$ valued connection on ${\cal M}$. These connections can be expressed in terms of the vielbein as usual. The holonomy condition means that the $Sp(1)_R$ curvature associated with the connection $\omega_{\alpha AB}$ vanishes. The vielbeins must furthermore obey the relations \cite{Bagger:1983tt}
\be
g_{\a\b} V^\a_{aA} V^\b_{bB} =\epsilon_{ab}\epsilon_{AB}\ ,
\qquad
V^\a_{aA} V^{\b aB} + \a \leftrightarrow \b = g^{\a\b} \d_A^B\ ,
\ee
where $\epsilon_{ab}$ and $\epsilon_{AB}$ are $Sp(n)$ and $Sp(1)_R$ invariant tensors. We use the conventions
\be
\zeta^a \e_{ab} = \zeta_b\ , \quad \epsilon^{ab} \zeta_b = \zeta^a\ , \quad \e^{ab} \e_{bc} = - \d_c^a
\ee
for raising and lowering indices with $\e_{ab}$ and similar conventions for $ \epsilon_{AB}$.
It is useful to define
\be
P_\mu^{aA}= \partial_\mu \phi^\alpha V_\alpha^{aA}\ .
\label{pdef}
\ee
We can write the globally supersymmetric boundary action for the hyperscalars as
\be
S_H^0 = \frac{1}{\tilde\lambda^2} \int d^6 x \bls - \fr14 P_\m^{a A} P_{aA}^\m  - \fr{i}{2} \bar{\z^a} \g^\m \cD_\m \z_a \brs\ ,
\ee
where $\cD_{\m} \z^a = \nabla_\mu\z^a +\partial_\mu \phi^\alpha \omega_\alpha^{ab} \z_b$, with $\nabla_\mu$ containing the Lorentz spin connection, and we have introduced a coupling constant $\tilde\lambda$. This action is invariant under the global supersymmetry transformations
\ba
\d \f^\a &= i \sq{2}  \bar{\e}^A \z^a V^\a_{aA}\ ,
\nn\w2
\d \z^a &= \fr{1}{\sq{2}}  \g^\m \e_A P_\m^{a A}\ .
\ea
We now consider the coupling of this boundary hypermultiplet action to our $D=7$ bulk supergravity system. We begin the construction by modifying the field transformations as
\ba
\d \f^\a &= i \sq{2} e^{-a\vf} \bar{\e}^A \z^a V^\a_{aA}\ ,
\nn\w2
\d \z^a &= \fr{1}{\sq{2}} e^{ a \vf} \g^\m \e_A P_\m^{a A}\ .
\ea
As before, we consider the boundary conditions that can be imposed on bulk fields such that these conditions form an orbit under supersymmetry. The bulk fermions on which we will attempt to impose non-zero boundary conditions transform under the projected supersymmetry as\footnote{As in \eqref{7DSUSYodd}, we have simplified the discussion by including only parity-odd fields which receive non-zero boundary conditions in these transformations.}
\be
\bs
\d \p_{\m - }^A &= - \fr12 K_{\m\n} \g^\n \e^A - \fr{i}{40} e^\vf F_{\r\s}^i \s^{iAB} (3 \g_\m \g^{\r \s} - 5 \g^{ \r\s} \g_\m ) \e_B \ , 
\\
\d \c_+^A &= - \fr14 e^{6 \a \f} \partial_{\usevenn} \hat{\s} \e^A - \fr{i}{20} e^\vf F_{\m\n}^i \s^{i AB} \g^{\m\n} \e_B\ .
\es
\ee
This means that the following set of boundary conditions form an orbit under supersymmetry:
\be
\bs
\p_{\m -}^A\ev &=  \fr{9}{10\sq{2}} b e^{ ( c- a) \vf} \z_a P_\m^{aA} - \fr{1}{10\sq{2}} b e^{ (c-a) \vf } \g_{\m\n} \z_a P^{\n aA}+ ({\rm fermi})^3\ ,
\\
\c_+\ev &= \fr1{10\sq{2}} b e^{ (c-a)} \vf \g^\m \z_a P_\m^{aA} + ({\rm fermi})^3\ ,
\\
e^{6 \a \f} \partial_{\usevenn} \hat{\s}\ev &=  \fr1{10} b e^{c \vf} P_{\m}^{aA} P^{\m}_{aA}
+ ({\rm fermi})^2\ ,
\\
F_{\m\n}^{i} \s^{iAB}\ev &= i b e ^{(c- 1)\vf} P_{[\m}^{aA} P_{\n]a}^B + ({\rm fermi})^2\ ,
\\
K_{\m\n}\ev &=  \fr12 b e^{c\vf} P_{\m}^{aA} P_{\n a A} - \fr1{20} b e^{c\vf} P_\r^{aA} P^\r_{aA}\, g_{\m \n} + ({\rm fermi})^2\ ,
\es
\ee
where $a$ , $b$ and $c$ are constants to be determined, and, as before, all other parity-odd fields in \eqref{odd} are set to zero at lowest order in fermions. Calculating the surface term produced upon variation of the bulk action under \eqref{7DSUSY} and then imposing these boundary conditions, we find the total non-invariance of the bulk supergravity action:
\be\label{varSGHyperBC}
\bs
\d S_{SG} + \d S^0_{GHY} +\d S^1_{GHY} &=  \frac{b}{\k^2} \int_{\pa M} d^6 x e  \blb  \frac{i}{24\sq{2}} e^{ (c-a)\vf - \s} \bar \e^A \g^\m \g^{\r\s\t} \g^{\n} \g_\m \z^a P_{\n aA} G_{\r\s\t}  \\
& - \fr{i}{2} e^{c\vf} \bar \e^A \g^{\m \n \r} \p_{\r B} P_{ \m a A} P_\n^{a B}  - \fr{i}{2} e^{c\vf} \bar \e^A \g^{\m \n } \p_{B} P_{ \m a A} P_\n^{a B} \\
& -\fr1{\sq{2}} e^{(c-a)\vf} \bar \e_A \s^{iAB} \z^a \cP_{\m}^i P^\m_{a B} \brb\ .
\es
\ee
Then, by the Noether procedure, we find the following boundary action
\be\label{bah}
\bs
S_H &= \frac{1}{\tilde\l^2} \int_{\pa M} d^6 x e \blb - \fr14 e^{2 a \vf} P_\m^{a A} P_{aA}^\m - \fr{i}{2} \bar{\z^a} \g^{\m} \cD_\m \z_a \\
&- \fr{i}{\sq{2}} e^{ a \vf} \bar \z^a \g^\m \g^\n \p_\m^A P_{\n aA} + i \sq{2} a e^{a\vf} \bar \z^a \g^\m \p^A P_{\m a A} \brb\ .
\es
\ee
Here we have set $c = 2a$ which is required for invariance. With this condition, the action varies to give
\be
\bs
\d S_H &= \frac{1}{\tilde\l^2} \int_{\pa M } d^6 x e \blb \fr{i}{\sq{2}} e^{a \vf} \bar \e^A \g^\n \g^\m (\cD_\m - D_\m) \left( \z^a P_{\n aA} \right) \\
& - \fr{i}{24 \sq{2} } e^{a\vf - \s} \bar \e^A \g^\m \g^{\r \s\t} \g^\n \g_\m \z^a P_{\n a A} G_{\r \s\t} \\
&+ \fr{i}{2} e^{2a \vf} \bar \e^A \g^{\m \n \r} \p_{\r B} P_{\m a A} P_\nu^{a B} + i a e^{2 a \vf} \e^A \g^{\m \n } \p_{B} P_{\m a A} P_\nu^{a B}\\
& + \fr1{2\sq{2}} e^{a \vf} \bar \e_A \s^{iAB}\left( 2 a \g^\mu \g^\nu + \g^\nu \g^\mu \right) \z^a \cP_{\m}^i P_{\n aB} \brb\ .
\es
\ee
The $D_\mu(\zeta P)$ term, with $D_\mu$ defined in \eq{covder} and \eq{PQQ}, arises from the variation of the $\zeta\psi_\mu P$ term. Furthermore, the ${\cal D}_\mu(\zeta P)$ term, with the covariant derivative defined with respect to the pull-backed connection $\partial_\mu\phi^\alpha \omega_{\alpha AB}$, comes from the variation of the $P^2$ term in \eq{bah}. The $PG, PP$  and ${\cal P}P$ terms  cancel the bulk surface term \eqref{varSGHyperBC}, as long as $ b = \fr{\k^2}{ \tilde\l^2 } $ and $ a= \fr12$, while the term proportional to $({\cal D}_\mu-D_\mu)(\zeta P)$ vanishes as long as the boundary $Sp(1)_R$ connection is set equal that for the bulk at the boundary location, {\it i.e.}\footnote{An analogous condition has been found in \cite{Gherghetta:2002nq} with all the bulk scalars set to zero.}
\be\label{conc}
Q_\m^{AB}\ev  = \pa_\m \f^\a \omega_\a^{AB}\ ,
\ee
where $Q_\mu^{AB} = \frac{i}{4} \epsilon^{ijk} Q_\mu^{jk} \sigma_i^{AB}$ and $Q_\mu^{jk}$ is defined in \eq{PQQ}.

Owing to the order in fermions to which we have been working, this equation  is valid only to purely bosonic order. We also note that the coupling of these boundary hypermultiplets does not produce any classical non-invariances such as those which arose for the vector multiplets.

Substituting \eqref{conc}  into the field strength for $Q_\m^{AB}$ and then using the boundary conditions $C|_{\partial M}=C^{i r}|_{\partial M} = 0$, we find
\be
P_{[\mu}^{a A} P_{\nu ] a}^B = - \fr{i}{4} \e^{i j k} \left(2 P_{[ \mu}^{ i r} P_{\nu ]}^{ j r}+ \fr{1}{2\sqrt{2}} \e^{ijl} C^{l r\pr} F_{\mu \nu}^{r\pr} \right) \sigma^{k A B} \ev\ .
\ee
This implies that the $Sp(1)_R$ curvature of the boundary hypermultiplets is identified with the $Sp(1)_R$ curvature of the bulk scalars. The fact that this is nonzero is consistent with the fact that the full manifold parametrised by the $4p+4$ scalars from the bulk and the $4m$ scalars from the boundary hypermultiplets parametrise a QKM in the limit of coincident boundaries.

As before, we note that a term of the form $\bar \z^a \g^{\mu \nu \rho} \z_a G_{\mu \nu \rho}$ is not present in the boundary action, although it is present in the 6D hypermultiplet coupled action  as given in Refs \cite{Nishino:1986dc, Nishino:1997ff} and in Ref.\ \cite{Gherghetta:2002nq}. At the purely bosonic order, as required for the coupling process considered in this section, the boundary condition simply sets $A_{\mu \nu \rho}$ equal to zero on the boundary. However, at higher order in fermions the boundary condition will be of the form $A_{ \mu \nu \rho}|_{\partial M} \sim \bar \z^a \g_{\mu \nu \rho} \z_a$. This will then give rise to the required term in the coincident boundaries limit in an analogous  way to that described in Section \ref{BoundaryAction} and Appendix A.

The scalar kinetic term in the boundary action \eqref{bah} is multiplied by an unusual factor $e^\vf$, which also results in the unusual Noether coupling  term $e^{\fr{\vf}{2}} \bar \z^a \g^\m \p^A P_{\m a A }$. This can be understood by bearing in mind that the hyperscalar $\vf$ as well as the newly-coupled boundary scalars must together form a QKM in the limit of coincident boundaries.

Note that the gauged $U(1)_R$ lies in the $SO(n,3)$ isometry group of the bulk sigma model. Furthermore, the boundary hyperk\"ahler manifold ${\cal M}$ does not necessarily have any isometries. Consequently, the gauge field $A_\mu^{r'}$ does not arise in the definition of the covariant derivative given in \eq{pdef}. However, the local $U(1)_R$ symmetry is nonetheless realised as a result of the the boundary condition \eq{conc}. This condition is crucial for the quaternionic K\"ahler structure on the overall scalar manifold, ${\cal N}$, which arises under local supersymmetry, as expected. The manifold ${\cal N}$ is a single irreducible QKM of dimension $4m+4p+4$, with coordinates $(\phi^\alpha, \phi^{ir'}, \Phi^I, \varphi)$, whose holonomy group is contained in $Sp(m+p+1)\times Sp(1)$. In the absence of the $m$ boundary hypermultiplets, and in the coincident boundaries limit, it is known that ${\cal N}$ can be described as the quaternionic K\"ahler coset $SO(p+1,4)/SO(p+1)\times SO(4)$ \cite{Bergshoeff:2005pq}.  In the presence of $m$ boundary hypermultiplets, however, the structure of the overall scalar manifold ${\cal N}$ arising in the coincident boundaries limit depends on the specific properties of ${\cal M}$. It would be interesting to determine, for example, the conditions on ${\cal M}$ under which ${\cal N}$ becomes a symmetric or homogeneous QKM.

\section{Extensions of the Model and Further Classical Anomalies}\label{extensions}
In order to cancel the complete set of anomalies, it is necessary to consider various modifications to the model described so far. One such modification is the addition of a bulk topological mass term for the $3$-form potential \cite{Townsend:1983kk,Bergshoeff:2005pq}. Another is the inclusion of further bulk Chern-Simons terms together with further modifications to the boundary conditions, while a third is the coupling of boundary-localised tensor multiplets. We will consider all three of these extensions in the following section.

\subsection{The Topological Mass Term}\label{top}

A topological mass term can be added to the bulk action described in Section \ref{sectwo}, thereby arriving at a one-parameter extension. However, a mass term of the form $h A_3 \wedge F_4$ with a constant mass parameter $h$ violates the $\ztwo$ symmetry of the boundary. In order to respect this $\ztwo$ symmetry, we need to allow the mass parameter $h$ to undergo a jump at the boundary location when viewed from an upstairs perspective. To accomplish this, we dualise $h$ to a $6$-form potential $A_6$ such that the field equation for $h$, now treated as a scalar field, equates $h$ to the dual of the $A_6$ field strength, while the field equation for $A_6$ implies that $h$ is at least piecewise constant. In this formulation, we can now assign odd parity to $h$ so as to render the term $h A_3 \wedge F_4$ parity-even. The resulting new terms in the bulk action are
\begin{equation}
S_h = \frac1{\kappa^2}\int_{M} d^7 x \hat{e} \bigg \lbrace - i h^2 e^{4 \hat{\sigma} } + h \hat{\varepsilon}^{MNRSTUV} \hat{G}_{MNRSTUV} \bigg \rbrace
\label{topaction}
\end{equation}
where
\begin{equation}
\begin{split}
\hat{G}_{MNRSTUV} &= 7 \partial_{[M} \hat{A}_{NRSTUV ] }  + \frac{1}{36} \hat{F}_{[MNRS} \hat{A}_{TUV]} - \frac{4 \sqrt{2} }{ 7! 3 } \hat{\varepsilon}_{MNRSTUV} e^{\frac32 \hat{\sigma} } C \\
& - \frac{i}{5!} e^{2 \hat \sigma} \hat{ \bar{\psi} }_{[M} \hat{\gamma}_{NRSTU} \hat{\psi}_{V ]}   + \frac{8i }{6!} e^{2 \hat \sigma} \hat{\bar\psi}_{[M} \hat{\gamma}_{NRSTUV]} \hat{\chi} \\
& + \frac{27i}{7!} e^{2 \hat \sigma} \hat{\bar\chi} \hat{\gamma}_{ MNRSTUV } \hat{\chi}  -  \frac{i}{7!} e^{2 \hat \sigma} \hat{\bar\lambda}^{\hat{r}} \hat\gamma_{ MNRSTUV} \hat{\lambda}
\end{split}
\end{equation}
and the new terms in the supersymmetry transformation rules are
\begin{equation}
\begin{split}
\delta \hat{\psi}_{M} &= - \frac{4}{5} h e^{2 \hat{\sigma} } \hat{\gamma}_M \hat{\epsilon}\ ,\\
\delta \hat{\chi} &= - \frac{16}{5} h e^{2 \hat{\sigma} } \hat{\epsilon}\ ,\\
\delta \hat{A}_{MNRSTU} &= - \frac{1}{63} \delta \hat{A}_{[MNR} \hat{A}_{STU]} + \frac{24i}{7!} e^{2 \hat{\sigma} } \hat{\bar\epsilon} \hat{\gamma}_{[MNRST} \hat{\psi}_{U]} - \frac{16 i}{7!} e^{ 2 \hat\s} \hat{\bar\epsilon} \hat{\gamma}_{MNRSTU} \hat{\chi}\ ,\\
\delta h &= 0\ .
\label{topsusy}
\end{split}
\end{equation}
The $6$-form potential $\hat{A}_{\mu \nu \rho \sigma \lambda \tau}$  is parity even and  $\hat{A}_{\mu \nu \rho \sigma \lambda 7}$ is parity odd.
The action is now invariant under a modified tensor gauge transformation under which $A_6$ must transform as
\begin{equation}
\begin{split}
\delta \hat{A}_{MNRSTU} &= - \frac{1}{21} \hat{A}_{[MNR} \pa_S \hat{\l}_{TU]}\ .
\label{deltaA6}
\end{split}
\end{equation}

In the presence of the boundaries, the supersymmetry of the bulk-plus-boundary action is unaffected by this construction and the variational principle remains consistent, provided that we impose the boundary condition
\be
h \ev =0\ ,\quad\quad A_{\mu_1...\mu_5 7}\ev =0\ .
\ee
However, we may also consider the boundary value of $h$ to be a constant
\be
h\ev = h_0\ .
\label{hzero}
\ee
This will lead to the introduction of a new boundary term and modified boundary conditions that will produce further classical anomalies in the boundary Yang-Mills gauge symmetry.

We now seek an orbit of boundary conditions which contains \eqref{hzero}. As we are interested in the effects of the topological mass term on classical anomalies, we consider boundary conditions involving boundary vector multiplets as well as the constant $h_0$. However, because the hypermultiplets do not effect the classical non-invariances, we will not further consider their simultaneous coupling here. Carrying out this process, we find an orbit of boundary conditions given by \eq{detBC} with the following modifications (up to quartic fermion terms):
\bea
e^{6 \alpha \phi} \partial_{\usevenn} \hat{\sigma}\ev &=&  -\frac{\kappa^2}{10\lambda^2} e^{-\sigma} H_{\mu \nu}^X H^{\mu \nu}_X - 2(4+\gamma) e^{\s + 2\vf} h_0\ ,
\nn\w2
K_{\mu\nu} \ev &=& \frac{\kappa^2}{2\lambda^2} e^{ - \sigma} H_{\mu \rho}^X H\indices{_\nu^\rho_X} - \frac{3\kappa^2}{40\lambda^2} e^{-\sigma} H_{\rho\sigma}^X H^{\rho\sigma }_X g_{\mu\nu} +\gamma e^{\s + 2\vf} h_0 g_{\m\n}\ ,
\nn\w2
C\ev &=& - \frac{30}{\sqrt 2} \left(\frac45 + \gamma\right) \frac{h_0}{g}\, e^{\vf+2\sigma}\ ,
\label{constBC}
\eea
where $\gamma$ is a parameter shortly to be determined.
To find the total supersymmetric action up to a supersymmetry anomaly, we need to give the total boundary action
\be
\bs
S_B^{\rm tot.} = \int_{\pa M} d^6 x & e \blb - \frac1{8\lambda^2} e^{- \sigma} H_{\mu\nu}^X H^{\mu\nu }_X  - \frac{i}{2\lambda^2} \bar \eta^X \gamma^\mu D_\mu \eta_X \\
&- \frac{i}{4\lambda^2} e^{-\frac\sigma2} H_{\rho\sigma}^X \bar\eta_X \gamma^\mu \gamma^{\rho\sigma} \psi_\mu - \frac{i}{4\lambda^2} e^{- \frac\sigma2} H_{\mu\nu}^X \bar\eta_X \gamma^{\mu\nu} \chi
\w2
& +\frac1{32\lambda^2 g^2} \varepsilon^{\mu\nu\rho\sigma\lambda \tau}  \omega_{\mu\nu\rho}^0 (C) \omega_{\sigma\lambda\tau}^0 (A)
\w2
&+\frac{4h_0}{\kappa^2} e^{\s+2\vf}  + \frac{7h_0}{\kappa^2}\varepsilon^{\m\n\r\s\l\t} A_{\m\n\r\s\l\t} \w2
&+\fr{ih_0 \k^2}{8\l^4} e^\s \o_{\m\n\r}^0(C) \bar \h^X \g^{\m\n\r} \h_X \brb\ .
\label{tba}
\es
\ee
Requiring supersymmetry up to a Wess-Zumino consistent anomaly determines the value of $\gamma$:
\be
\gamma=-\frac45\ .
\ee
It is interesting that this implies the boundary condition $C\ev=0$.
One can further check that the above boundary conditions are consistent with the variational principle. The variation of the action \eqref{tba} under tensor gauge transformations subject to the boundary conditions \eqref{detBC} gives the additional gauge anomaly contribution
\be
\delta_\Lambda S_B^{\rm tot} = - \fr{h_0 \k^2}{8 \l^4} \int_{\pa M}d^6x\, e \varepsilon^{\m\n\r\s\l\t}  H_{\m\n}^X H_{\r \s X} \pa_\l C_\t^Y \L_Y\ .
\ee

Correspondingly, there is an additional contribution to the supersymmetry anomaly given by
\be
 - \fr{h_0 \k^2}{8 \l^4} \int_{\pa M}d^6x\, e \varepsilon^{\m\n\r\s\l\t} \blb  H_{\m\n}^X H_{\r \s X} \d_\e C_\l^Y  C_{\t Y}  - 2 \o_{\m\n\r}^0(C)  H^X_{\s\l} \d_\e C_{\t X}\brb\ .
\ee
As before, one may check that the inclusion of these anomalies continues to give a Wess-Zumino consistent system.

\subsection{Additional Bulk Chern-Simons Terms, Boundary Conditions and Classical Anomalies}\label{CS}

Before evaluating  the gauge/Lorentz anomalies that result from the variation of the bulk plus boundary action subject to the chosen boundary conditions, we need to discuss possible additional extensions of the bulk model. Terms of types that may produce anomalous variations are  of the forms $A_3 \wedge \tr\, R\wedge R, \ \omega_{7L},\ \omega_7(A),\ \omega_3 (A)\wedge \tr\, R\wedge R$ where $\omega_{7L}$ and $\omega_7 (A)$ are the Lorentz and Yang-Mills Chern-Simons forms, respectively.\footnote{While a term of the type $\omega_7 (A)$ does arise in the $SO(5)$ gauged maximal $D=7$ supergravity, it does not appear in any gauged half-maximal $D=7$ supergravity. The half-maximal truncation of the maximal theory studied in \cite{Cvetic:2003xr} might seem to indicate the presence of $\omega_7(A)$  but, in fact, such a term is not allowed by supersymmetry in this system.}  The $\omega_{7L}$ and $A_{(3)}\,\tr\, \wedge R\wedge R$ terms are known to arise in the $K3$ compactification of $D=11$ supergravity supplemented with the Duff-Minasian term $A_{(3)}\, \tr\, R\wedge R\wedge R\wedge R$. These have been used in a Ho\v{r}ava-Witten formulation of \emph{ungauged} pure $D=7$ supergravity \cite{Gherghetta:2002nq}. However, in the non-compact $D=7$ model we are considering here,
derivation from higher dimensions involves a noncompact internal space of infinite volume. Indeed, as we saw in the Introduction, a 3-manifold of this kind, known as $H(2,2)$, is involved in the reduction from $N=1$, $D=10$ supergravity to the $SO(2,2)$ gauged supergravity in $D=7$ \cite{Cvetic:2003xr}, yielding a consistent Kaluza-Klein truncation. The same model can also be obtained from $D=11$ supergravity by reducing on $H(2,2)\times S_1$, again yielding a consistent Kaluza-Klein truncation. However, in the presence of the term $D=11$ $A_{(3)}\, \tr\, R\wedge R\wedge R\wedge R$ and even in the presence of the Yang-Mills sector in $D=10$, a consistent Kaluza-Klein ansatz is not at present known. A preliminary investigation of the infinite volume problem\footnote{We would like to acknowledge detailed discussions with Chris Pope on this point.} suggests that the appropriate Weyl rescaling of fields needed to obtain finite kinetic terms in $D=7$ leads to vanishing coefficients in front of the $\omega_7(A)$ term and we expect this to be the case for the $\omega_3(A) \tr\, R\wedge R$ term as well.  With this in mind, we shall not consider further the inclusion of higher-derivative terms in the bulk Lagrangian as given in Section \ref{sectwo}, but supplemented by the topological mass term added in Section \ref{top}. However, we shall consider modifications of the boundary condition on $A_{(3)}$ occasioned by the inclusion of Chern-Simons terms for the bulk gauge fields and Lorentz connection such that
\be
A_{(3)}^{\rm extra} \ev = c_A\omega_3(A)+c_L\omega_{3L} + {\rm (fermi)}^2 \ ,
\label{mod}
\ee
where $c_A$ and $c_L$ are arbitrary constant coefficients. Extending the full set of supersymmetric boundary conditions \eq{detBC} to incorporate this modification will, in particular, alter the boundary condition on the extrinsic curvature $K_{\mu\nu}$ which will now must include terms taking the form
\be
K_{\mu\nu} \ev \sim e^{-\s} F_\mu{}^{\rho r'} F_{\nu\rho}^{r'} + e^{-\s} R_\mu{}^{\rho\, \um\un} R_{\nu\rho\,\um\un} +\cdots\ .
\ee
Since $K_{\mu\nu}$ picks up contributions for the boundary stress tensor, it follows that modifications proportional to this, in turn, imply that the full boundary action must contain terms given by
\be
S_B^{\rm ext.} \sim \int_{\partial M} d^6 x e \blb e^{-\s} F_{\mu\nu}^{r'} F^{\mu\nu r'} + e^{-\s} R_{\mu\nu\,\um\un} R^{\mu\nu\,\um\un} +\cdots \brb \ .
\label{mossr2}
\ee
An $R^2$ term of this type has been encountered in the Ho\v{r}ava-Witten formulation of $D=11$ supergravity compactified on $S^1/\ztwo$ \cite{Moss:2008ng}. We note that the dilaton factors in \eqref{mossr2} are equivalent to the dilaton factor multiplying the kinetic term in \eqref{NoetherCoupledVectorAct}. In standard $D=6$ calculations, higher-derivative invariants with either $e^{\sigma}$ or $e^{-\sigma}$ factors multiplying the $R^2$ term are possible \cite{Bergshoeff:1986wc,Lu:2010ct}. Supersymmetrizing the $e^{\sigma}$ variant would imply the presence of a term of the form $ B_2 \wedge R_2 \wedge R_2 $, whilst supersymmetrizing the $e^{-\sigma} $ variant implies that the 3-form field strength appearing in the action is Chern-Simons modified such that $ G_3 = dB_2 + \omega_{3L}$.  Since the boundary condition \eqref{mod} implies that the field strength becomes Chern-Simons modified in the coincident boundaries limit (see Appendix \ref{coincident}) we deduce that the necessary factor here must be $e^{-\sigma}$ multiplying the $R^2$ term present in this boundary action. A similar argument also applies to vector couplings, which is consistent with the fact that Noether coupling forced us to determine the coefficient $a=-1$ in Section \ref{BoundaryAction}.

To summarise, the total action we have constructed so far is the sum of \eq{7DAction}, \eq{topaction}, \eq{tba} and \eq{mossr2}. In this action, the following terms contribute to the bosonic anomaly:
\be
-\frac1{2{\sqrt 2}\kappa^2 g^2} \int_M d^7 x {\hat A}_{(3)} \wedge {\hat F}^{\hat r} \wedge {\hat F}^{\hat r} + 7! \int_{\partial M} d^6 x h_0 A_{(6)}\ .
\label{relevant}
\ee
Using the modified boundary conditions \eqref{mod}, the variations of these terms give the new total bosonic anomaly
\be
\bs
\Omega_6^1 =& \int_{\pa M} \blb \fr{ 2
h_0}{\k^2}\Big( \big( \fr{2 c_A}{3} - \fr1{8g^2 h_0} \big)  \o_2^1(A) + \fr{ 2 c_L}{3} \o_{2L}^1 +
\fr{\k^2}{2 \l^2} \o_2^1(C) \Big) \wedge \\
& \Big( \big( \fr{2 c_A}{3} - \fr1{8g^2 h_0} \big)  \tr F \wedge F + \fr{ 2 c_L}{3}R \wedge R +
\fr{\k^2}{2 \l^2} \tr H_2 \wedge H_2 \Big) \\
& - \fr{1}{32 \k^2 g^4 h_0} \o_2^1(A) \wedge \tr F\wedge F \brb \ ,
\es
\label{AnomMinusTens}
\ee
where\footnote{Note that we are using the Chern-Simons 3-form normalisation given in Equation \eqref{fs}, as in Reference \cite{Bergshoeff:2005pq}, for both gauge and Lorentz symmetries. This gives rise to the factors of $\frac13$ in the descent relations.} $\omega_2^1$ is defined by $\delta\omega_3^0=\frac13 d\omega_2^1$.
If we consider the gauge group for the boundary vector multiplets $K^\prime$  to be the tensor product of simple groups $K_1 \otimes \ldots \otimes K_{n_g} $, we can define the 4-forms  $ G^\ua $, where $\ua =0, \ldots , n_g +1 $, as 
\be
G^{\uzero}= \tr F\wedge F ,\quad G^{\uone}= \tr R \wedge R, \quad  G^{\utwo} =  \tr H^{(1)} _2 \wedge H_2^{(1)},  \ \ldots  \ , G^{\underline{n_g +1}} = \tr H_{2}^{(n)} \wedge H_{2}^{(n)} ,
\ee
where $d\omega_3^0(A) = \frac13 \tr F\wedge F = \frac13 F^{r'} \wedge F^{r'}$ and $d\omega_3^0(C) = \frac13 \tr H_2\wedge H_2 = \frac13 H_2^X\wedge H_2^X$. Then the anomaly (8.16) is related to the following $8$-form polynomial
\bea
\Omega_8^{clas} &=& \frac{8h_0}{\kappa^2} \bls ( \frac13 c_A G^{\uzero} + \frac13 c_L G^{\uone} + \sum_{\ua = 2}^{n_g+1}  \frac{\kappa^2}{4(\lambda^\ua)^2} G^\ua \brs  \wedge
\\
&& \bls \left(\frac13 c_A-\frac{1}{8h_0 g^2}\right) G^{\uzero}+\frac13 c_L G^{\uone}
+ \sum_{\ua = 2}^{n_g+1}  \frac{\kappa^2}{4(\lambda^\ua)^2} G^\ua \brs \ ,
\nn\label{clasanomaly}
\eea
by the descent equations $\omega_8^{clas}=d\Omega_7^0$ and $\delta\Omega_7^0 =d\Omega_6^1$.

\subsection{Boundary Tensor Multiplets and Further Classical Anomalies}\label{tensors}
The classical non-invariance produced so far obeys the Wess-Zumino consistency conditions and produces terms of the correct forms to cancel the quantum anomalies. However the classical anomaly produced is still not sufficiently general to completely cancel the anomalies produced by quantum effects and so to yield an overall invariant system. We therefore consider a further extension of the model by adding $n_T$ boundary-localised tensor multiplets to the action. These multiplets have the form $(B_{\m\n}^x,\c_-^{Ax}, \f^x)$, where $x = 2, \ldots , n_T+1$, which play a crucial role in the implementation of a generalized Green-Schwarz anomaly cancellation mechanism introduced in \cite{Sagnotti:1992qw}.

Tensor multiplets of this form are known to exist in rigid $D=6$ supersymmetry and accordingly a coupling process similar to that shown in Sections \ref{BCs} and \ref{BoundaryAction} will be possible. However this process is complicated by the fact that the 3-form field strength $H^x_3=dB^x_2$ is required, by closure of the supersymmetry algebra, to be self-dual: $H_3 = \star H_3$. This has the consequence that the na\"{\i}ve kinetic term that one would write for $B^x_2$ vanishes. This problem may be addressed by use of a non-manifestly Lorentz invariant action \cite{Schwarz:1993vs}, or by reformulating the problem at the equation-of-motion level. We shall not attempt here a full analysis of these couplings. Although a full coupling would be necessary for detailed analysis of the classical supersymmetry anomalies, it is not necessary for analysis of the purely bosonic anomalies. This is due to the fact that bosonic anomaly contributions arising from boundary tensors can only be generated by the variation of one type of term in the boundary action. This crucial anomaly-generating term type is analogous to the bulk Chern-Simons term $\fr1{g^2} A_3 \wedge F_2^{\hat r}  \wedge F_2^{\hat r}$, and is of the same form as the standard anomaly counterterm that is seen in purely $D=6$ theories \cite{RicSag}. In our boundary action, it appears as
\be
\int_{\pa M} v_{\ua}^x B_2^x \wedge G^\ua_4 \ ,
\label{TensorTerm}
\ee
where $v_\ua^x$ is a numerical coupling matrix analogous to the $\fr{1}{g^2}$ which appears in the in the bulk action, and where summation over the index $x=2,...,n_T+1$ is understood. If $B^x$ is required to transform under the bosonic symmetries of the theory according to
\be
\d B^x = v^{\pr x}_{\ua} \o_2^{1\ua} \ ,
\ee
then the variation of \eqref{TensorTerm} will produce a non-invariance of the form
\be
\int_{\pa M}   v^x_{\ua} v_{ \underline{b} }^{ \pr x} \o_2^{1\ua} G^\ub \ .
\ee
Adding this to the classical anomaly generated so far, we can write the total anomaly as
\be
\Omega_6^{1 tot} = \int_{\pa M}  v^\uI_{\ua} v_{\ub}^{ \pr \uJ}  \eta_{\uI \uJ} \o_2^{1\ua} \wedge G^\ub \ ,
\ee
where the index $x$ has been extended to a new index $\uI =  0, \dots , n_T+1$. In general, the index $\ua=0,...,n_g+1$. However, if $n_g < n_T$, then the matrix $v^\uI_{\ua} v_{\ub}^{ \pr \uJ}  \eta_{\uI \uJ}$ has non-maximal rank, which turns out to put a severe restriction on the quantum anomaly polynomial \cite{Sagnotti:1992qw,Gherghetta:2002nq}. This restriction is lifted for $n_T\ge n_g$. For simplicity, we shall assume that $n_T=n_g$ from here on. Then, we find that the vector $v^\uI_\ua$ is given by
\be
\bs
v^{\uzero}_\ua &= v^{\pr \uzero}_\ua =\Big( \fr{ 2  c_A \sqrt{ - 2 h_0} }{3 \k} - \fr{1}{4 g^2 \sqrt{ - 2 h_0}  }, \ \fr{2  c_L \sqrt{ - 2 h_0}  }{3\k},\ \fr{\k \sqrt{- 2 h_0} }{2 (\l^{\utwo})^2}, \ldots,\ \fr{\k \sqrt{ - 2 h_0} }{2 (\l^{\underline{n_g+1}})^2} \Big) \ ,  \w4
v^{\uone}_\ua & =  v^{\pr \uone}_\ua = \Big(  \fr{1}{4  \k g^2 \sqrt{ - 2 h_0} }, 0, 0, \ldots ,  0 \Big) \ , \quad \quad
v^\uI_\ua = v^{\underline x}_\ua , \quad v^{ \pr \uI}_{\ua}  = v^{\pr \underline x} _\ua , \ \text{for} \ \uI  = 2, \ldots, n_T +1  \ ,
\es
\ee
 $\eta_{\uI \uJ} = \text{diag}( -, +, \ldots , + )$ and we have assumed $h_0 < 0$ which makes the components of these vectors real. This represents the full classical anomaly which will be cancelled against the quantum anomalies to be described in the next section.

\section{Quantum Anomalies and Anomaly Cancellation} \label{anomalies}

We shall now construct an example of an anomaly-free model in the $D=7/D=6$ Ho\v{r}ava-Witten setting that we have been constructing in this paper. As we wish to end up with an $R$-symmetry gauged model, we need to start with a matter-coupled noncompact gauged $D=7$ theory. The possible non-compact gauge groups and the surviving even-parity bulk fields have been listed in \cite{Bergshoeff:2005pq}. Here, we shall consider the $SO(2,1)$ gauged $D=7$ model which consists of minimal supergravity coupled to one vector multiplet. The bulk scalars parametrize the coset $SO(1,3)/SO(3)$ and the $SO(1,2)$ subgroup of $SO(1,3)$ is gauged. The structure constants are given by \cite{Bergshoeff:2005pq}
\be
{\hat f}_{\hat I\hat J\hat K}=\e_{\underline{ijk}}\ ,\quad
\underline i= 1,2,4\ ,
\ee
where $\e_{\underline{ijk}}$ are the $SO(1,2)$ structure constants.
In \eq{indexsplit}, we now have $p=0, n=1$, and the resulting even-parity fields form the multiplets
\be
(e_\mu^{\um}, \psi_{\mu+}, B_{\mu\nu}^-)\ ,\qquad (B_{\mu\nu}^+, \chi_-,\sigma)\ , \qquad (\psi_-,\varphi,\Phi_I)\ , \qquad (A_\m^4, \l_+^4 )\ ,
\ee
with supersymmetry transformations as given in \eq{6DSUSY}.
The vector field $A_\mu^4$ gauges the R-symmetry group $U(1)_R$. We have denoted the $D=6$ chiralities of the fermions explicitly for convenience, and we have split the $2$-form potential into parts that have self-dual and anti-self-dual field strengths.


The chiral fermions $(\psi_{\mu +}, \chi_-, \lambda_+^4, \psi_-)$ give rise to gravitational, $U(1)_R$ and mixed gravitational-$U(1)_R$ anomalies on the boundaries. The anomalies are encoded in an $8$-form polynomial made up of the Riemann and Yang-Mills curvature forms, via the descent equations. The standard anomaly formulae give

\bea
 \Omega(  \psi_{\m +} ) &=& \frac{5}{24} F_1^4 - \frac{19}{96} F_1^2
~trR^2 +\frac{1}{5760}~\left[245~\tr~R^4
               -\frac{5\times 43}{4}~(\tr~R^2)^2 \right] \ ,
\nn\w2
\Omega(\chi_-) &=& -\frac{1}{24}~F_1^4 - \frac{1}{96} F_1^2 ~trR^2
    -\frac{1}{5760}~\left[\tr~R^4 +\frac54~(\tr
R^2)^2\right]\ ,
\nn\w2
\Omega(\l_+^4) &=& \frac{1}{24}~F_1^4 + \frac{1}{96} F_1^2 ~trR^2
    +\frac{1}{5760}~\left[\tr~R^4 +\frac54~(\tr
R^2)^2\right]\ ,
\nn\w2
\Omega(\psi_-) &=&
    -\frac{1}{5760}~\left[\tr~R^4 +\frac54~(\tr R^2)^2\right]\ ,
\nn\w2
\Omega( B_{\mu\nu +}) &=& \frac{1}{5760}\left[ -28~\tr~R^4
+10~(\tr~R^2)^2\right]\ ,\label{p4}
\eea
where $F_1$ is the $U(1)_R$ field strength, and we have suppressed the wedge symbol, so that, for example $F_1^2 \tr R^2= F_1\wedge F_1 \wedge \tr R\wedge R$.

The total anomaly coming from the bulk fields on each boundary is half of the total bulk anomaly. Thus on a given boundary we have
\be \Omega^{bulk}_{grav/U(1)_R}|_{\partial M_1}=  \frac{5}{48} F_1^4 -\frac{19}{192} F_1^2 ~trR^2
+\frac{1}{5760}~ \left[122~\tr~R^4 -\frac{55}{2}~(\tr~R^2)^2 \right]\
\label{abulk} .
\ee
Next, we need to compute the quantum anomalies that result from the introduction of $n_V$ gauge, $n_H$ hyper and $n_T$ tensor multiplets on a given boundary. It is useful first to compute the total gravitational anomaly. Summing up the bulk contributions given in \eq{abulk} and those of the boundary multiplets, the total gravitational anomaly on $\partial M_1$ is given by
\bea
\Omega^{\textit tot.}_{grav.}|_{\partial M_1} &=&  \ft{1}{5760} \Big[ (n_V-n_H-29n_T+122) \tr\, R^4 \nn\\
&& \ \ + \ft54 (n_V-n_H+7 n_T -22)\, (\tr\, R^2)^2 \Big]\ .
\label{grav}
\eea
The $\tr\,R^4$ term must necessarily vanish for anomaly freedom. As we have assumed that there is no bulk Lorentz Chern-Simons term, the vanishing of the $\tr\,R^4$ anomaly imposes the constraint \footnote{
In the standard $N=1$, $D=6$ anomaly cancellation, the equivalent relation is given by $n_H-n_V+29n_T=273$. The difference here is due to two factors. Firstly, our $n_T$ counts the number of boundary-localised tensor multiplets whilst the $n_T$ in the standard equation counts the total number of tensor multiplets. As one tensor multiplet comes from the reduction of the bulk supergravity multiplet, our $n_T$ differs from the standard setup by $1$. Secondly, the quantum anomaly in our case is split across two boundaries and so differs from the standard result by a factor of 2. Therefore in our case we have a different gravitational-anomaly cancellation condition from the standard condition: $n_H-n_V+29n_T = (273 - 29)/2 = 122$.}
\be
n_H-n_V+29n_T=122\ . \label{r4}
\ee
Using this condition in \eq{grav}, and including the contributions to the $U(1)_R$ and mixed gravitational-$U(1)_R$ anomalies (\ie the $F_1^4$ and $F_1^2 \tr\,R^2$ terms in \eq{abulk}, together with similar contributions from all the boundary matter multiplets that have been introduced), we find
\bea
 \Omega^{\textit tot.}_{grav/U(1)_R}|_{\partial M_1} &=&  \ft{1}{128} (n_T-4) (\tr\, R^2)^2 +\ft1{48} \left[ 2(n_V-n_T) +5\right] F_1^4
\nn\w2
&&+\ft1{192} \left[ 2(n_V-n_T)-19 \right] F_1^2 \tr\, R^2 \ .
\label{grav2}
\eea
At this point, we need to specify $n_V$, $n_H$ and $n_T$ such that the condition \eq{r4} is satisfied, where the boundary Yang-Mills gauge group has total dimension $n_V$, and such that the $n_H$ hyperfermions form a set of representations of this group.  A complete analysis of all the possibilities is beyond the scope of the present paper. Instead, we shall give one example to illustrate how anomaly freedom can be achieved in the bulk-plus-boundary system that we have constructed. We shall take the gauge group on a given boundary to be 
\be
K'= E_6\times E_7\ ,
\ee
so that $n_V=78+133$. Furthermore, we shall introduce two tensor multiplets, and five hypermultiplets in fundamental representations of $E_6$ and five fundamental representations of $E_7$. Thus, all in all, we have
\bea
n_T &=& 2\ ,
\nn\w2
n_V &=& 78+133\ ,
\nn\w2
n_H &=& 5\times (27,1) + 5\times (1,56)\ .
\label{data1}
\eea
Using this data and employing the relations
\be
\bs
 \Tr\, H_6^2= 4 \tr\, H_6^2\ ,\quad \Tr\, H_6^4= \ft12 (\tr\, H_6^2)^2\ , \quad \tr\, H_6^4 = \ft{1}{12}(\tr\, H_6^2)^2\ ,
\w2
\Tr\, H_7^2= 3 \tr\, H_7^2\ ,\quad \Tr\, H_7^4= \ft16 (\tr\, H_7^2)^2\ ,\quad \tr\, H_7^4 = \ft{1}{24} (\tr\, H_7^2)^2\ ,
\label{anomaly2}
\es
\ee
where $\Tr (\tr)$ denote the trace in the adjoint (fundamental) representation, we find that the total one-loop anomaly polynomial is encoded by
\bea
\Omega_8^{1-loop} &=& -\frac1{64}\left(\tr\,R^2\right)^2 +\frac{141}{16} F_1^4 +\frac{133}{64} F_1^2 \tr\, R^2
\nn\w2
&& +F_1^2 \left( \tr\, H_6^2 +\frac34 \tr\, H_7^2 \right) -\frac1{96} \tr\, R^2 \left( \tr\, H_6^2 +2 \tr\, H_7^2\right)
\nn\w2
&& +\frac1{576} \left[ 2 \left(\tr\, H_6^2\right)^2- \left(\tr\, H_7^2\right)^2 \right]\ .
\label{oneloop}
\eea
%


Now we shall require that this quantum anomaly polynomial cancels the classic anomaly polynomial \eq{clasanomaly} with $n_T=n_g=2$. We begin by making the following redefinitions
\ba
\tilde \l^{\uone} &= \l^{\uone}  \left( \fr{1}{ - h_0 \k^2  }\right)^{\fr14} & \tilde \l^{\utwo} &= \l^{\utwo}  \left( \fr{1}{ - h_0 \k^2  }\right)^{\fr14}& \tilde g &= g \left(  - h_0 \k^2 \right)^\fr14 \nn\\
\tilde c_A &= c_A \left( \fr{ - h_0}{\k^2} \right)^{\fr12} & \tilde c_L &= c_L \left( \fr{ - h_0}{\k^2} \right)^{\fr12}\ ,
\ea
where all the new parameters are dimensionless. This allows us to rewrite the anomaly polynomial \eq{clasanomaly} as
\bea
\Omega_8^{clas} &=& -8 \left( \frac13 \tilde c_A G^{\uzero} + \frac13 \tilde c_L G^{\uone} + \frac{1}{4(\tilde \l^1)^2} G^{\utwo} +\frac{1}{4 (\tilde \l^2) ^2} G^{\uthree} \right) \wedge
\nn \\
&& \left( \left(\frac13 \tilde c_A + \frac{1}{8 \tilde g^2}\right)G^{\uzero} +\frac13 \tilde c_L G^{\uone}
 + \frac{1}{4(\tilde \l^1)^2} G^{\utwo} +\frac{1}{4 (\tilde \l^2) ^2} G^{\uthree} \right) \nn \\
&&+ v_{\ua}^2 v_{\ub}^{\pr 2}  G^\ua \wedge G^\ub+ v_{\ua}^3 v_{\ub}^{\pr 3} G^\ua \wedge G^\ub\ .
\label{Oclas}
\eea
In order for the system to be anomaly free, \eqref{Oclas} must cancel the quantum anomaly polynomial
\bea
\Omega_8^{1-loop} &=&  \frac{141}{16} (G^{\uzero})^2  -\frac1{64} (G^{\uone})^2+\frac{133}{64} G^{\uzero} G^{\uone}
\nn\w2
&& +G^{\uzero} \left( G^{\utwo} +\frac34 G^{\uthree} \right) -\frac1{96} G^{\uone} \left( G^{\utwo} +2 G^{\uthree} \right)
\nn \w2
&& +\frac1{576} \left[ 2 (G^{\utwo})^2- (G^{\uthree})^2 \right ]\ .
\label{oneloopreex}
\eea
This requirement places 10 constraints on the 21 parameters in \eqref{Oclas} which leaves an 11 dimensional space of solutions. In order to demonstrate that a solution exists in which all parameters are real, we give an example solution\footnote{
Finding solutions to a large number of simultaneous equations such as these is greatly simplified by finding the Groebner basis for the equations. This is most easily done using the program Singular or the Mathematica package STRINGVACUA.},
\ba
\tilde c_A &=  0.0000 &  \tilde g & = 0.1443 & \tilde c_L &= 0.0000 &  \tilde \l^1 &= 3.4641 & \tilde \l^2 &= 4.0000 \nn \\
v^2_0 &= 0.0000, & v^2_1 &= -3.6424 & v^2_2 &=1.4106 &  v^2_3 &= -1.0000 \nn\\
v^{\pr 2}_0 &= 0.0000 &v^{\pr 2}_1 &= -0.0074 & v^{\pr 2}_2 &=  0.0000 & v^{\pr 2}_3 &=  -0.0037 \nn \\
v^3_0 &=  -1.0000 &  v^3_1 &= -0.2303 &v^3_2 &=  0.0000 & v^3_3 &=0.0000 \nn\\
v^{\pr 3}_0  &= 8.8125 & v^{\pr 3}_1 &= 0.0490 &  v^{\pr 3}_2 &=  0.0000 & v^{\pr 3}_3 &= 0.0000 \ ,
 \ea
where we have dropped the underlines in $v^\uI_{\ua}$ for notational simplicity.
This demonstrates that anomaly-free bulk-plus-boundary models can indeed be constructed as we have described.

\section{Conclusions}\label{conclusion}

We may view the construction in this paper as a worked example of an anomaly-free model with gauged R-symmetry and a positive cosmological potential. A variety of approaches has been followed in the search for realistic reductions of string/M-theory to candidate effective $D=4$ theories. The standard compactifications and brane constructions limit to effective supergravity theories which populate only a sub-class of the available models that one might want to explore, however. In particular, the class of non-compact gaugings of supergravity has been rather under-exploited to date. Such models depart from models with compact gauged R-symmetries, such as the original $D=4$ gauged $N=8$ supergravity \cite{de_Wit:1982ig}. The discovery of models with gauged R-symmetries then led on to searches for models with gauged non-compact symmetry groups \cite{Gunaydin:1985cu,Hull:1984wa}. These were in turn obtained by reduction from higher dimensions on non-compact manifolds \cite{Hull:1988jw}.

The physical interest of models with non-compact gaugings is illustrated by cosmological approaches such as the SLED program of supersymmetry in large extra dimensions \cite{Burgess:2007ui}, which takes as a starting-point example the $D=6$ Salam-Sezgin model \cite{Salam:1984cj}. But non-compact gaugings have not yet figured prominently in the search for realistic string or M-theory particle physics vacua. One reason for this has been the lack of a perceived link to the ``ur-theories'' in $D=10$ and $D=11$. A path towards such links has now been opened up, however, by the reduction in Ref.\ \cite{Cvetic:2003xr}, involving precisely the sort of non-compact manifold reduction envisaged in \cite{Hull:1988jw}. So, it seems that a relevant chapter in the encyclopedia of string/M-theory reductions has only just been opened.

In the present paper, we have focused primarily on a process for generating a chiral, anomaly-free model starting from a gauged R-symmetry In order to provide a richer and more fully worked-out scheme for $D=6$ models such as those needed for the SLED program, we began with a gauged R-symmetry model in $D=7$. To generate a chiral theory in $D=6$, we used a Ho\v{r}ava-Witten construction based on a slice of $D=7$ bulk spacetime bounded by two $D=6$ spaces which can then be populated with $D=6$ supermatter as needed to construct an anomaly-free model. Ho\v{r}ava-Witten type constructions, generalising the original $D=11$/$D=10$ construction of the heterotic string from M-theory \cite{Horava:1996ma,Horava:1995qa}, can also be seen as domain-wall brane-solution constructions such as the $D=5$/$D=4$ ``heterotic M-theory'' construction \cite{Lukas:1998yy,Lukas:1998tt}. These naturally produce chiral theories in the lower even dimension. But this then raises the issue of potential quantum anomalies in the reduced theory. The mechanism of anomaly cancellation involves anomaly inflow from the bulk higher-dimensional space together with a careful choice of ``matter'' fields to populate the boundary brane spaces. In the $D=11$/$D=10$ construction, this uniquely yields the original $\mathrm{E}_8$ gauge multiplet on each bounding brane \cite{Horava:1996ma,Horava:1995qa,Moss:2004ck,Moss:2005zw,Moss:2008ng}. As one goes down in dimensionality, the anomaly-cancellation requirements become less stringent, so that in a direct $D=5$/$D=4$ analysis \cite{Lukas:1999nh}, the only anomalies requiring cancellation are gauge and mixed gravitational-gauge anomalies, with a wide resulting set of anomaly-free constructions. The present $D=7$/$D=6$ construction presents an intermediate scenario, with a detailed set of cancellation requirements as presented in Section \ref{anomalies}. These do not uniquely specify the boundary gauge groups and fields, but they do impose a stringent set of anomaly-cancellation conditions on them. In the present paper, we have not attempted a comprehensive study of the solutions to these conditions, but it may be hoped that such a study might reveal classes of phenomenologically interesting scenarios.

The main challenges to be met in carrying out the $D=7$/$D=6$ construction revolved around the details of coupling 8-supercharge boundary matter to the 16-supercharge bulk theory. One needs to take care to provide necessary Gibbons-Hawking-York terms so as to ensure consistency between the bulk-plus-boundary variational equations and the chosen boundary conditions for the bulk fields. The halving of the supersymmetry at a boundary is a natural consequence of any Ho\v{r}ava-Witten type orbifold construction. But one needs to take great care here in handling the supersymmetric couplings, since in the absence of a fully off-shell formalism, the classical boundary non-gauge-invariances of the bulk theory, as needed for anomaly inflow, engender also supersymmetry anomalies.

The occurrence of supersymmetry anomalies in Ho\v{r}ava-Witten type constructions is already familiar from the work of Refs \cite{Moss:2004ck, Moss:2005zw}, but what is different about the constructions made in the present paper is the order at which these occur. In \cite{Moss:2004ck, Moss:2005zw}, an iterative construction to suppress the anomalies was carried out in powers of the boundary coupling constant for the original $D=11/D=10$ heterotic construction. In that case, the $D=10$ boundary action and the corresponding boundary conditions for $D=11$ bulk fields occurred at first order in the boundary coupling $\fr1{\l^2}$ 
\ba
S_{\text{boundary}} & \sim \fr1{\l^2} \int_{\pa M} * F_\tw \we F_\tw & C_\th \ev \sim  \fr{\k^2}{\l^2} \o_\th
\ea
The bosonic anomaly, however, comes from substituting the boundary condition for $C_\th$ into the variation of the Chern-Simons term, 
\ba
\d S \sim \fr1{\k^2} \int_{\pa M} \d C_\th \we C_\th \we F_\fo \sim \fr1{\k^2}  \bigg( \fr{k^2}{\l^2} \bigg)^3 \int_{\pa M} \d \o_\th \we \o_\th \we F_\tw \we F_\tw 
\ea
which gives an anomaly at third order in $\fr1{\l^2}$. This means that supersymmetric Noether coupling can be caried out to second order in $\fr1{\l^2}$ \cite{Moss:2004ck} without interference from anomaly complications, whose discussion can be postponed until later on at third order in $\fr1{\l^2}$ \cite{Moss:2005zw}. In the construction of the present paper, however, the discussion of anomalies cannot similarly be postponed. This is because the bosonic anomaly in this case comes from a variation 
\ba
\d S \sim \fr1{\k^2} \int_{\pa M} \d A_\th \we \o_\th(A) \sim \fr1{\k^2}  \fr{k^2}{\l^2} \int_{\pa M} \d \o_\th (C) \we \o_\th (A)
\ea
which occurs already at first order in $\fr1{\l^2}$, \ie it is of the same order as the boundary action that we are constructing. 

Thus, the best that one can arrange for in the present bulk-plus-boundary coupling is agreement with the Wess-Zumino consistency conditions, as discussed in Section \ref{BoundaryAction}. Reduction of the $D=7$/$D=6$ construction to a purely $D=6$ theory by taking a coincident boundary limit, as explained in Appendix \ref{coincident}, confirms the correctness of this construction by yielding precisely the $D=6$ Wess-Zumino consistent system that was found in Ref.\ \cite{RicSag}. It is interesting to note that the construction of supersymmetric bulk-plus-boundary systems, similar to those considered here, is greatly simplified by the use of the \emph{`susy without b.c.'} formalism considered in  \cite{Belyaev:2007bg}. This formalism, as currently constructed, requires an off-shell supersymmetry realisation and so works only in cases with lesser degrees of supersymmetry. However, in the future this may provide a deeper understanding of complicated constructions such as those made in this paper.

Another challenge encountered in the present construction is the coupling of boundary hypermultiplets. These are in general necessary in order to arrange for gravitational anomaly cancellation, but they do not affect the classical gauge or supersymmetry anomalies. However, the bulk-plus-boundary couplings in this sector lead to novel problems. Eight-supercharge ($N=2$, $D=4$ or $N=1$, $D=6$ supersymmetry) hypermultiplets coupled to supergravity require an overall quaternionic K\"ahler target-space manifold \cite{Bagger:1983tt}. Indeed, the bulk $D=7$ theory dimensionally reduced to $D=6$ and truncated to $N=1$, $D=6$ local supersymmetry generates precisely this kind of scalar target-space manifold \cite{Bergshoeff:2005pq}. However, when one includes additional hypermultiplets on the $D=6$ boundaries of the Ho\v{r}ava-Witten construction, one runs into the problem that one cannot simply add quaternionic K\"ahler manifolds to produce an overall quaternionic K\"ahler manifold. The resolution of this problem led to the connection condition \eqref{conc}.

A number of aspects of the constructions discussed in this paper call for further development. A fuller treatment of the hypermultiplet couplings will be given in a separate publication, and a full analysis of the solutions to the anomaly-cancellation conditions is called for. Another open question deals with a very special class of remarkably anomaly-free
$D=6$ theories with gauged $U(1)_R$ symmetries. These are:
\begin{itemize}
\item the $E_7\times E_6 \times U(1)_R$ invariant model in which
the hyperfermions are in the $(912,1,1)$ representation of the gauge
group \cite{RandjbarDaemi:1985wc},
\item the $E_7\times G_2\times U(1)_R$ invariant model with
hyperfermions in the $(56,14,1)$ representation of the gauge group
\cite{Avramis:2005qt}, and
\item the $F_4\times Sp(9)\times U(1)_R$ invariant model with
hyperfermions in the $(52,18,1)$ representation of the gauge group
\cite{Avramis:2005hc}.
\end{itemize}
We have determined that the construction of this paper cannot yield any of these models in a coincident brane limit. Thus, finding the higher-dimensional origins of these theories, if any, remains an outstanding open problem.

More generally, the r\^ole of noncompact gaugings and their higher-dimensional origins through reduction on noncompact spaces needs further consideration. Noncompact reductions may, as in the $H(2,2)$ reduction considered in \cite{Cvetic:2003xr}, yield classically consistent Kaluza-Klein reductions. But at the quantum level, this classical Kaluza-Klein consistency is surely broken. Moreover, noncompact reductions from higher-dimensional theories would be expected to lead to a continuous Laplace eigenvalue spectrum without a mass gap between the retained lower-dimensional and the higher truncated Kaluza-Klein states. One can imagine a number of possible responses to this situation. One would be to consider a compactification of the reduction space, perhaps by modding out by discrete symmetries, but this would also likely be at the cost of introducing supersymmetry breaking at some new scale in the problem. Another might be to look for discrete Laplace eigenfunctions in the midst of a continuous-eigenvalue spectrum. Such situations are not unusual in other contexts, such as condensed-matter physics. It remains to be seen whether they have a relevance in the context of noncompact gauged R-symmetries.

\section*{Acknowledgements}

We would like to acknowledge collaboration with Chris Pope and Eric Bergshoeff in early stages of this project and for many subsequent discussions. We also thank Alex Kehagias for useful discussions. For hospitality during the course of the work, ES would like to thank the Theoretical Physics Group at Imperial College London and National Technical University of Athens; KSS and TGP would like to thank the George P. and Cynthia Woods Mitchell Institute for Fundamental Physics and Astronomy at Texas A\&M University; KSS would also like to thank the TH Unit at CERN and the Albert Einstein Institute, Potsdam. KSS would like to thank as well the Mitchell Family for hospitality and a beautiful and quiet place to work during the Cooks Branch workshop in April 2010. TGP would like to thank Noppadol Mekareeya for many helpful discussions. The research of E.S. was supported in part by NSF grants PHY-0555575 and PHY-0906222. The work of K.S.S. was supported in part by by the STFC under rolling grant PP/D0744X/1.


\begin{appendices}

\section{The Coincident Boundary Limit}\label{coincident}

We now consider taking the coincident boundaries limit when the boundaries are populated with vector multiplets as described in Section \ref{BoundaryAction}. This gives a six-dimensional gauged supergravity theory similar to that described in \cite{Bergshoeff:2005pq}.

The orbit of boundary conditions in this $D=7$ system involves both Neumann and Dirichlet types, which have different effects on the reduced system. Let us first consider the Neumann boundary conditions with the example of the form field $A_{\m \n \r}$. This is subject to two boundary conditions: one on the $x^7=0$ boundary and the other on the $x^7=L$ boundary (where $L$ is the interval length ). We can follow the work of \cite{Moss:2004ck,LukOvWald1,LukOvWald2} and use the fact that, in the limit of small interval length, it is sufficient to approximate the value of $A_{\mu \nu \rho}$ in the bulk by a linear interpolation between the two boundary conditions:
\be
A_{\m \n \r} = A_{\m \n \r} \bigg |_{ x^7=0} \left( 1 - \fr{ x^7 }{L} \right) + A_{\m \n \r} \bigg |_{x^7 = L} \fr{ x^7}{L} \ .
\ee
We consider the simplified case in which the boundary at $x^7=0$ is populated by vector multiplets in the way we have described and the boundary at $x^7=L$ is empty. This means that the bulk field $A_{\mu \nu \rho}$ becomes
\be
A_{\m \n \r} = \left( \frac{3\kappa^2}{4\lambda^2} \omega^0_{ \mu\nu\rho} (C)  + \frac{i\kappa^2}{8\lambda^2} e^{ \sigma} \bar\eta^X \gamma_{\mu\nu\rho} \eta^X \right) \left( 1 - \fr{x^7}{L} \right) \ .
\ee
This causes the six-dimensional 3-form field strength to become Chern-Simons modified:
\be
\bs
\hat F_{\m \n \r 7} &=  3 \pa_{[\m} \hat A_{\n \r] 7} - \pa_{7} \hat A_{\m\n\r} \\
&= \sq{2} \left( 3 \pa_{[\m} B_{\n \r]} + \frac{3}{2g'^2} \o_{\m \n\r}^0(C) + \frac{i}{4g'^2} \bar \h^X \g_{\m\n\r} \h_X \right) \ ,
\es
\ee
where we have defined ${g\pr}^2 = \fr{2 L \l^2}{\k^2}$ in order to match the conventional result. If we now redefine $G_{\m\n\r}$ as the appropriately normalised bosonic part in the above equation \ie
\be
G_{\m\n\r} = 3 \pa_{[\m} B_{\n \r]} + \fr{3}{{2g\pr}^2} \o_{\m \n\r}^0(C) \ ,
\ee
then we find that $G_{\m\n\r}$ is invariant under the Yang-Mills gauge symmetry since $B_{\m\n}$ develops a gauge transformation due to the boundary condition \eqref{LambdaBC}:
\begin{equation}
\delta_{\Lambda} B_{\mu \nu} = - \frac{1}{{g\pr}^2} \partial_{[\mu} C_{\nu]}^{X} \Lambda_{X} \ .
\end{equation}
On the other hand, the field $\hat \s$ receives a Dirichlet boundary condition. In the small interval limit, we can again interpolate between its two boundary values such that
\be
\pa_7 \hat \s = \pa_7 \hat \s \bigg |_{ x^7=0} \left( 1 - \fr{ x^7 }{L} \right) + \pa_7 \hat \s \bigg |_{x^7 = L} \fr{ x^7}{L} \ .
\ee
If we integrate this equation and impose the requirement that the average value of $\hat \s$ is the same as in the empty boundaries case, then we obtain
\be
\hat \s = - \pa_7 \hat \s \bigg |_{x^7 = 0} \left( \fr{(x^7)^2}{2L} - x^7 + \fr{L}{3} \right) + \pa_7 \hat \s \bigg |_{x^7 = L} \left( \fr{(x^7)^2}{2L} - \fr{L}{6} \right) + \fr45 \s + \fr25 \vf \ .
\ee
Performing similar steps for all fields that receive non-trivial boundary conditions and then incorporating these into the $D=7$ bulk action together with the Gibbons-Hawking-York terms and the boundary action, and ignoring any higher-order terms in $\fr{1}{\l^2}$ or $L$, we obtain the $D=6$ action
\be
\bs
S_{SG (6)} =& \frac{2L}{{\kappa}^2} \int  dx^6 e \bigg \lbrace \frac14 {R} - \frac1{8g^2} e^{{\sigma}}
{F}_{\mu \nu}^{r^\prime} {F}^{\mu \nu r^\prime } - \fr{1}{8{g\pr}^2} e^{-\s} H_{\m \n}^X H^{\m\n}_X - \frac{1}{12} e^{-2 {\sigma}} {G}_{\mu\nu\rho} {G}^{\mu\nu\rho} - \frac14 \partial_\mu {\sigma} \partial^\mu {\sigma}
\\
& - \frac14 \partial_\mu {\varphi} \partial^\mu {\varphi}  - \frac14 P_\mu^{i {r}} P^{\mu i {r}} - \frac14 \mathcal{P}_\mu^{{r}} \mathcal{P}^{\mu {r}} - \frac14 \mathcal{P}_\mu^{i } \mathcal{P}^{\mu {i}}- \frac18 g^2 e^{-\sigma} \left( C^{i {r^\prime}} C^{i {r^\prime}} + 2S^{i {r^\prime}} S^{i {r^\prime}} \right)
\\
& + \frac{1}{16g^2}  \varepsilon^{\mu\nu\rho\sigma\lambda\tau} B_{\m\n} F_{\r\s}^{r\pr} F_{\l\t}^{r\pr} +  \fr{1}{32g^2{g\pr}^2}  \ve^{\m\n\r\s\l\t} \o_{\m\n\r}^0 (C) \o_{\s\l\t}^0 (A) \\
& -\frac{i}{2} {\bar{\psi}}_\mu {\gamma}^{\mu\nu\rho} {D}_\nu {\psi}_\rho - \frac{i}{2}  {\bar{ \chi}} {\gamma}^\mu {D}_\mu \chi - \frac{i}{2g^2} {\bar{\lambda}}^{{r^\prime}} {\gamma}^\mu {D}_\mu {\lambda}_{{r^\prime}}
\\
&- \frac{i}{2}  {\bar{ \psi}} {\gamma}^\mu {D}_\mu \psi- \frac{i}{2}  {\bar{ \psi}}^{r} {\gamma}^\mu {D}_\mu \psi^{r} - \fr{i}{2{g\pr}^2} \bar\h^X \g^\m D_\m \h_X
\\
& - \frac12 {\bar{\psi}}^{{r}} \sigma^i {\gamma}^\mu {\gamma}^\nu {\psi}_\mu P_\nu^{i {r}} - \frac12 {\bar{\psi}} \sigma^i {\gamma}^\mu {\gamma}^\nu {\psi}_\mu \mathcal{P}_\nu^{i}- \frac{i}{2}  {\bar{\psi}}^{{r}} {\gamma}^\mu {\gamma}^\nu {\psi}_\mu \mathcal{P}_\nu^{ {r}}
\\
&- \frac{i}{2} {\bar{\chi}} {\gamma}^\mu {\gamma}^\nu {\psi}_\mu \partial_\nu {\sigma}  - \frac{i}{2} {\bar{\psi}} {\gamma}^\mu {\gamma}^\nu {\psi}_\mu \partial_\nu {\varphi} - \frac{i}{24 } e^{{-\sigma}} {G}_{\mu\nu\rho} \bigg (\bar\psi_{[\lambda} \gamma^\lambda \gamma^{\mu\nu\rho} \gamma^\tau \psi_{\tau]}
\\
& -2 \bar \psi_\lambda \gamma^{\mu\nu\rho} \gamma^\lambda \chi - \bar\chi \gamma^{\mu\nu\rho} \chi + \bar\psi \gamma^{\mu \nu \rho} \psi + \bar\psi^r \gamma^{\mu\nu\rho} \psi^r - \frac1{g^2} \bar\lambda^{r^\prime} \gamma^{\mu\nu\rho} \lambda^{r^\prime} - \fr{1}{{g\pr}^2} \h^X\g^{\m\n\r} \h_X \bigg)
\\
& - \frac14 \mathcal{P}_\mu^i \bigg( {\bar\psi}_{[\rho} \sigma^i  \gamma^{\rho} \gamma^{\mu} {\gamma}^\tau \psi_{\tau]} + {\bar\chi} \sigma^i \gamma^{\mu} {\chi} + \frac1{g^2} {\bar\lambda}^{{r}^\prime} \sigma^i \gamma^{\mu} \lambda^{{r}^\prime}
\\
& + \fr{1}{{g\pr}^2} \bar \h^X \s^i \g^\m \h_X -{\bar\psi}^{{r}} \sigma^i \gamma^{\mu} \psi^{{r}} - {\bar\psi} \sigma^i \gamma^{\mu} \psi \bigg)
\\
& - \frac{i}{4g^2} e^{\frac{{\sigma}}{2}} {F}_{\mu\nu}^{{r^\prime}} \bigg( {\bar\psi}_\rho \gamma^{\mu\nu} \gamma^{\rho} \lambda^{{r^\prime}} + {\bar\chi} \gamma^{\mu\nu} \lambda^{{r}^\prime} \bigg) - \fr{i}{4{g\pr}^2} e^{-\fr\s2} H_{\m\n}^X \bigg( \bar \p_\r \g^{\m\n} \g^\r \h_X - \bar \c \g^{\m \n} \h_X \bigg)
\\
& - i \mathcal{P}_\mu^r \bar\psi \gamma^\mu \psi^r - e^{\frac{\sigma}{2}} C^{i r r^\prime} \bar\lambda^{r^\prime} \sigma^i \psi^r + i e^{\frac{\sigma}{2}} S^{r r^\prime} \bar\lambda^{r^\prime} \psi^r - e^{\frac{\sigma}{2}} S^{i r^\prime} \bar\lambda^{r^\prime} \sigma^i \psi
\\
&+ \frac{1}{2 \sqrt{2}} e^{-\frac{{\sigma}}{2}} \lambda^{r^\prime} \sigma^i \gamma^\mu \psi_\mu \left( C^{i {r^\prime}} - \sqrt{2} S^{i r^\prime} \right) + \frac{1}{2 \sqrt{2}} e^{-\frac{{\sigma}}{2}} \lambda^{r^\prime} \sigma^i \chi \left( C^{i {r^\prime}} - \sqrt{2} S^{i r^\prime} \right)\bigg \rbrace\ .
\label{6DAction2}
\es
\ee
Carrying out the reduction of the supersymmetry transformations and averaging over $x^7$ gives
\begin{equation}
\begin{split}
\delta {e}_\mu^{\um} &= i {\bar\epsilon} \gamma^{\um} \psi_\mu\ \ ,\\
\delta {\psi}_\mu &=  {D}_\mu \epsilon + \frac{1}{24} e^{-\sigma} G_{\rho\sigma\tau}  \gamma^{\rho\sigma\tau} \gamma_\mu \epsilon - \frac{i}{2} \mathcal{P}_\mu^i \sigma^i \epsilon\ , \\
\delta \chi &= - \frac12 \gamma^\mu \partial_\mu \sigma  \epsilon - \frac{1}{12} e^{-\sigma} {G}_{\mu\nu\rho} \gamma^{\mu\nu\rho} \epsilon\ ,\\
\delta {B}_{\mu\nu} &= - i e^{\sigma} \bar\epsilon \gamma_{[\mu} \psi_{\nu]} + \frac{i}2 e^\sigma \bar\epsilon \gamma_{\mu\nu} \chi\ +  \fr{1}{{g\pr}^2}  \d_\e C_{[\m}^X C_{\n] X}\ ,\\
\delta \sigma &= - i {\bar\epsilon} \chi\ ,\\
\delta {A}_\mu^{{r^\prime}} &=  i e^{-\frac{\sigma}{2}} {\bar{\epsilon}} \gamma_\mu \lambda^{{r^\prime}}\ ,\\
\delta \lambda^{r^\prime} &= -\frac14 e^{\frac{\sigma}{2}} \gamma^{\mu \nu} F^{r^\prime}_{\mu \nu} \epsilon - \frac{i}{2 \sqrt{2}} g^2  e^{-\frac{\sigma}{2}} \left( C^{i r^\prime} - \sqrt{2} S^{i r^\prime} \right) \sigma^i \epsilon\ ,\\
\delta \psi &= \frac{i}{2} \gamma^\mu \left( \mathcal{P}_\mu^i \sigma^i - i \partial_\mu \varphi \right) \epsilon\ ,\\
\delta \psi^r &= \frac{i}{2} \gamma^\mu \left( P_\mu^{i r} \sigma^i + i \mathcal{P}_\mu^r \right) \epsilon\ , \\
\delta \varphi &= i \bar{\epsilon} \psi\ ,\\
\delta L_I^r &= \bar{\epsilon} \sigma^i \psi^r L_I^i\ ,\\
\delta L_I^i &= \bar{\epsilon} \sigma^i \psi^r L_I^r\ ,\\
\delta \Phi^I &= - L^{Ii} e^{-\varphi} \bar{\epsilon} \sigma^i \psi - i L^{Ir} e^{-\varphi} \bar{\epsilon} \psi^r\ ,\\
\delta C_\mu^X &= i e^{  \frac{\sigma}{2}}  \bar \epsilon \gamma_\mu \eta^X\ ,\\
\delta \eta^X &= - \frac14 e^{- \frac\sigma2}  \gamma^{\mu \nu} H_{\mu \nu}^X \epsilon\ .
\end{split}
\end{equation}
Under these supersymmetry transformations, the action varies into the supersymmetry anomaly
\begin{equation}
\begin{split}
\delta_{\epsilon} S &=  \frac{2L}{32\k^2 g^2 {g\pr}^2} \int_{\partial M} d^6 x e \bigg \lbrace \epsilon^{\mu\nu\rho\sigma \lambda \tau} H_{\mu \nu}^X H_{\rho\sigma}^X \delta_\epsilon A_\lambda^{r^\prime} A_\tau^{r^\prime}
- 2 \epsilon^{\mu\nu\rho\sigma \lambda \tau} \omega_{\mu\nu\rho}^0 (C) F_{\sigma \lambda}^{r^\prime} \delta_\epsilon A_\tau^{r^\prime} \\
& + \epsilon^{\mu\nu\rho\sigma \lambda \tau} F_{\mu \nu}^{r^\prime} F_{\rho\sigma}^{r^\prime} \delta_\epsilon C_\lambda^X C_\tau^X
- 2 \epsilon^{\mu\nu\rho\sigma \lambda \tau} \omega_{\mu\nu\rho}^0 (A) H_{\sigma \lambda}^X \delta_\epsilon C_\tau^X \bigg \rbrace\ \ ,
\end{split}
\label{AnomClasSUSY}
\end{equation}
which is Wess-Zumino consistent with its gauge variation,
\begin{equation}
\delta_\Lambda S =  \frac{2L}{32\k^2} \frac1{g^2 {g\pr}^2}\int_{\partial M} d^6 x e \bigg \lbrace  \varepsilon^{\mu \nu\rho \sigma \lambda \tau} H_{ \mu \nu}^X H_{\rho \sigma}^X \partial_\lambda A_\tau^{r^\prime} \Lambda^{r^\prime} + \varepsilon^{\mu \nu\rho \sigma \lambda \tau} F_{ \mu \nu}^{r^\prime} F_{\rho \sigma}^{r^\prime} \partial_\lambda C_\tau^{X} \Lambda^{X} \bigg \rbrace\ .
\label{AnomClasGauge}
\end{equation}
We note that the action and variations obtained here are consistent with the general matter coupled $D=6$ supergravity described in \cite{Nishino:1997ff, RicSag} for the case of a single tensor multiplet.

We note also that that if one were to consider the boundary matter coupling starting from the  boundary condition $A_{\mu \nu \rho} \sim c_A \omega^0_{\mu\nu\rho} (A)$ as described in Section \ref{CS}, then the reduced action would appear to contain kinetic terms of the form
\be
 S \sim \int d^6x e \left(  - \fr{1}{g^2}  e^\s  - c_A e^{-\s} \right) F_{\mu \nu}^{r\pr} F^{\mu \nu r\pr}
\ee
which is known to exhibit interesting phase transition behaviour \cite{Duff:1996cf, Duff:1996rs}. The dilaton dependence arises from supersymmetry considerations as described in Section \ref{BoundaryAction}.

\section{ $D=7$ 2-Form Supergravity}

We now consider the equivalent construction for the theory in which the 3-form $\hat A_{MNR}$ has been dualised into a 2-form $\hat B_{MN}$. This has the $D=7$ bulk action
\begin{equation}
\begin{split}
S_{SG} =& \frac{1}{\kappa^2} \int  d^7x \hat{e} \bigg \lbrace \frac12 \hat{R}(\hat\Gamma) - \frac14 \frac{1}{g^2} e^{\hat{\sigma}}
\hat{F}_{MN}^i \hat{F}^{MN i } - \frac1{4g^2}  e^{\hat{\sigma}} \hat{F}_{MN}^{\hat{r}} \hat{F}^{MN \hat{r}} - \frac{1}{12} e^{2 \hat{\sigma}} \hat{G}_{MNR} \hat{G}^{MNR}
\\
& - \frac58 \hat{\partial}_M \hat{\sigma} \hat{\partial}^M \hat{\sigma} -\frac12 \hat{P}_M^{i \hat{r}} \hat{P}^{M i \hat{r}} - \frac14 g^2 e^{-\sigma} \left( C^{i \hat{r}} C^{i \hat{r}} - \frac19 C^2 \right) -\frac{i}{2} \hat{\bar{\psi}}_M \hat{\gamma}^{MNR} \hat{D}_N \hat{\psi}_R
\\
& - \frac{5i}{2}  \hat{\bar{ \chi}} \hat{\gamma}^M \hat{D}_M \hat\chi - \frac{i}{2g^2}  \hat{\bar{\lambda}}^{\hat{r}} \hat{\gamma}^M \hat{D}_M \hat{\lambda}_{\hat{r}} - \frac{5i}{4} \hat{\bar{\chi}} \hat{\gamma}^M \hat{\gamma}^N \hat{\psi}_M \hat{\partial}_N \hat{\sigma} - \frac{1}{2g} \hat{\bar{\lambda}}^{\hat{r}} \sigma^i \hat{\gamma}^M \hat{\gamma}^N \hat{\psi}_M P_N^{i \hat{r}}
\\
& + \frac{i}{24 \sqrt{2}} e^{\hat{\sigma}} \hat{G}_{MNR} \bigg ( \hat{\bar\psi}_{[L} \hat\gamma^L \hat\gamma^{MNR} \hat\gamma^T \hat\psi_{T]} + 4 \hat{\bar\psi}_L \hat{\gamma}^{MNR} \hat\gamma^L \hat\chi - 3 \hat{\bar\chi} \hat\gamma^{MNR} \hat\chi + \frac{1}{g^2} \hat{\bar\lambda}^{\hat{r}} \hat\gamma^{MNR} \hat{\lambda}^{\hat{r}} \bigg)
\\
& + \frac{1}{8g} e^{\frac{\hat\sigma}2} \hat{F}_{MN}^i \bigg( \hat{\bar\psi}_{[L} \sigma^i  \hat\gamma^{L} \hat\gamma^{MN} \hat{\gamma}^T \hat\psi_{T]} - 2 \hat{\bar\psi}_L \sigma^i \hat\gamma^{MN} \hat\gamma^L \hat\chi + 3 \hat{\bar\chi} \sigma^i \hat\gamma^{MN} \hat{\chi} - \frac{1}{g^2} \hat{\bar\lambda}^{\hat{r}} \sigma^i \hat\gamma^{MN} \hat\lambda^{\hat{r}} \bigg)
\\
& - \frac{i}{4g^2}  e^{\frac{\hat{\sigma}}{2}} \hat{F}_{MN}^{\hat{r}} \bigg( \hat{\bar\psi}_L \hat\gamma^{MN} \hat\gamma^{L} \hat\lambda^{\hat{r}} + 2 \hat{\bar\chi} \hat\gamma^{MN} \hat\lambda^{\hat{r}} \bigg) + \frac{1}{2 \sqrt{2}} e^{-\frac{\hat{\sigma}}{2}} C^{i \hat{r}} \bigg( \hat{\bar\psi}_M \sigma^i \hat\gamma^M \hat\lambda^{\hat{r}} - 2 \hat{\bar\chi} \sigma^i \hat\lambda^{\hat{r}} \bigg)
\\
&- \frac{i\sqrt{2}}{24} g e^{-\frac{\hat\sigma}{2}}  C \bigg( \hat{\bar\psi}_M \hat\gamma^{MN} \hat\psi_N +2 \hat{\bar\psi}_M \hat\gamma^M \hat\chi + 3 \hat{\bar\chi} \hat{\chi} - \frac{1}{g^2} \hat{\bar\lambda}^{\hat{r}} \hat{\lambda}^{\hat{r}} \bigg) + \frac{1}{2g} e^{-\frac{\hat\sigma}{2}} C^{\hat{r} \hat{s} i} \hat{\bar\lambda}^{\hat{r}} \sigma^i \hat\lambda^{\hat{r}}  \bigg \rbrace
\label{7DActiontwo}
\end{split}
\end{equation}
where
\begin{equation}
\hat{G}_{MNR} = 3 \partial_{[M} \hat{B}_{NR]} - \frac{3}{\sqrt{2}g^2} {\hat\omega}^0_{MNR} (\hat{A})
\end{equation}
and all other definitions remain the same as before. This action has no Chern-Simons term, so we might expect no anomaly to occur. However, as we now see, this is not the case.

The action is invariant under the following local supersymmetry transformations:
\begin{equation}
\begin{split}
\delta \hat{e}_M^{\uM} &= i \hat{\bar\epsilon} \gamma^{\uM} \hat\psi_M\ ,\\
\delta \hat{\psi}_M &=  2 \hat{D}_M \hat\epsilon - \frac{1}{60 \sqrt{2}} e^{\hat\sigma} \hat{G}_{RST} \left( \hat\gamma_M \hat\gamma^{RST} + 5 \hat\gamma^{RST} \hat\gamma_M \right) \hat\epsilon \nonumber \\
&-\frac{i}{20g} e^{\frac{\hat\sigma}{2}}  \hat{F}_{RS}^i \sigma^i \left( 3 \hat\gamma_M \hat\gamma^{RS} - 5 \hat\gamma^{RS} \hat\gamma_M \right) \hat\epsilon - \frac{\sqrt{2}}{30} g e^{-\frac{\hat\sigma}{2}}  C \hat\gamma_M \hat{\epsilon}\ ,
\\
\delta \hat\chi &= - \frac12 \hat\gamma^M \hat{\nabla}_M \hat\sigma  \hat\epsilon - \frac{i}{10} e^{\frac{\hat\sigma}{2}} \hat{F}_{MN}^i \sigma^i \hat\gamma^{MN} \hat{\epsilon} - \frac{1}{15 \sqrt{2}} e^{\hat\sigma} \hat{G}_{MNR} \hat\gamma^{MNR} \hat\epsilon + \frac{\sqrt{2}}{30} e^{-\frac{\hat\sigma}{2}} C \hat\epsilon\ ,
\\
\delta \hat{B}_{MN} &= i \sqrt{2} e^{-\hat{\sigma}} \left( \hat{\bar\epsilon} \hat\gamma_{[M}\hat\psi_{N]} + \hat{\bar\epsilon} \gamma_{MN} \hat\chi \right) - \sqrt{2} \frac{1}{g^2} \hat{A}^{\hat{I}}_{[M} \delta \hat{A}_{N] \hat{I}}\ ,
\\
\delta \hat{A}_M^{\hat{I}} &= - g e^{\frac{\hat\sigma}{2}} \left( \hat{\bar\epsilon} \sigma^i \hat\psi_M + \hat{\bar\epsilon} \hat\gamma_{MN} \hat\chi \right) L^{\hat{I} i} + i e^{-\frac{\hat\sigma}{2}} \hat{\bar{\epsilon}} \hat\gamma_M \hat\lambda^{\hat{r}} L^{\hat{I} \hat{r}},
\\
\delta \hat\sigma &= - 2 i \hat{\bar\epsilon} \hat\chi\ ,
\\
\delta L_{\hat{I}}^i &= \frac{1}{g}\hat{\bar\epsilon} \sigma^i \hat\lambda^{\hat{r}} L_{\hat{I}}^{\hat{r}}\ ,
\\
\delta L_{\hat{I}}^{\hat{r}} &= \frac{1}{g}\hat{\bar\epsilon} \sigma^i \hat{\lambda}^{\hat{r}} L_{\hat{I}}^i \ ,
\\
\delta \hat\lambda^{\hat{r}} &= -\frac12 e^{\frac{\hat\sigma}{2}} \hat{F}_{MN}^{\hat{r}} \hat\gamma^{MN} \hat\epsilon + i g \hat\gamma^M \hat{P}_M^{i \hat{r}} \sigma^i \hat\epsilon - \frac{i}{\sqrt{2}} g e^{-\frac{\hat\sigma}{2}} C^{i \hat{r}} \sigma^i \hat\epsilon\ ,
\end{split}
\label{7DSUSYtwo}
\end{equation}
as well as having a $\ztwo$ symmetry which acts as before but now with $\hat B_{\mu \nu}$ assigned even parity and $\hat B_{\mu 7}$ odd parity. The action possesses a gauge symmetry under which $\hat B_{MN}$ transforms as
\begin{equation}
\delta_{\Lambda} \hat{B}_{MN} =  \frac{\sqrt{2}}{g^2} \hat \partial_{[M} \hat A^{\hat{I}}_{N]} \Lambda_{\hat{I}} \ .
\end{equation}

Once again, we begin our construction on a manifold with boundary by adding Gibbons-Hawking-York terms
\begin{equation}
\begin{split}
S_{GHY} &= \int_{\partial M} d^6 x \sqrt{-\hat{h}} \bigg \lbrace
 \hat{K} -\frac{i}{4} \hat{\bar\psi}_\mu \hat{\gamma}^{\mu \nu} \hat\psi_\nu  - \frac{5i}{4} \hat{\bar\chi} \hat\chi  \bigg \rbrace\ .
\end{split}
\label{defGHtwo}
\end{equation}
Redefining exactly as before but now with $B_{\mu \nu} = \frac{1}{\sqrt{2}} \hat{B}_{\mu \nu} $, $ G_{\mu \nu 7} = \frac{1}{\sqrt{2}} \hat G_{\mu \nu 7 } $ gives the $D=6$ supergravity transformations \cite{Bergshoeff:2005pq}
\begin{equation}
\begin{split}
\delta {e}_\mu^{\um} &= i {\bar\epsilon} \gamma^{\um} \psi_\mu\ ,\\
\delta {\psi}_\mu &=  {D}_\mu \epsilon - \frac{1}{24} e^{\sigma} G_{\rho\sigma\tau}  \gamma^{\rho\sigma\tau} \gamma_\mu \epsilon - \frac{i}{2} \mathcal{P}_\mu^i \sigma^i \epsilon\ , \\
\delta \chi &= - \frac12 \gamma^\mu \nabla_\mu \sigma  \epsilon - \frac{1}{12} e^{\sigma} {G}_{\mu\nu\rho} \gamma^{\mu\nu\rho} \epsilon\ ,\\
\delta {B}_{\mu\nu} &= i e^{-{\sigma}} \left( {\bar\epsilon} \gamma_{[\mu}\psi_{\nu]} + \frac12 {\bar\epsilon} \gamma_{\mu\nu} \chi \right) - \frac{1}{g^2} {A}^{{r^\prime}}_{[\mu} \delta {A}_{\nu] {r^\prime}}\ ,\\
\delta \sigma &= - i {\bar\epsilon} \chi\ ,\\
\delta {A}_\mu^{{r^\prime}} &=  i e^{-\frac{\sigma}{2}} {\bar{\epsilon}} \gamma_\mu \lambda^{{r^\prime}} \ ,\\
\delta \lambda^{r^\prime} &= -\frac14 e^{\frac{\sigma}{2}} \gamma^{\mu \nu} F^{r^\prime}_{\mu \nu} \epsilon - \frac{i}{2 \sqrt{2}} g^2 e^{-\frac{\sigma}{2}} \left( C^{i r^\prime} - \sqrt{2} S^{i r^\prime} \right) \sigma^i \epsilon \ ,
\\
\delta \psi &= \frac{i}{2} \gamma^\mu \left( \mathcal{P}_\mu^i \sigma^i - i \nabla_\mu \varphi \right) \epsilon \ ,\\
\delta \psi^r &= \frac{i}{2} \gamma^\mu \left( P_\mu^{i r} \sigma^i + i P_\mu^r \right) \epsilon\ , \\
\delta \varphi &= i \bar{\epsilon} \psi\ ,\\
\delta L_I^r &=  \bar{\epsilon} \sigma^i \psi^r L_I^i \ ,\\
\delta L_I^i &=  \bar{\epsilon} \sigma^i \psi^r L_I^r \ ,\\
\delta \Phi^I &= -L^{Ii} e^{-\varphi} \bar{\epsilon} \sigma^i \psi - i L^{Ir} e^{-\varphi} \bar{\epsilon} \psi^r\ ,
\end{split}
\end{equation}
where now $G_{\mu \nu \rho} = 3 \partial_{[\mu} B_{\nu \rho]} - \frac{3}{2g^2} \omega_{\mu \nu \rho}^0 ( A) $ and $B_{\mu \nu}$ transforms as
\begin{equation}
\delta_{\Lambda} B_{\mu \nu} = \frac{1}{g^2} \partial_{[\mu} A_{\nu]}^{r^\prime} \Lambda^{r^\prime}\ .
\end{equation}

Again, we can construct a consistent set of boundary conditions and in this we case find,\footnote{Here we have, as in the previous case, set all the free parameters that can occur equal to values that will be required by the variational principle, anticipating the final constructed boundary action.}
\begin{equation}
\begin{split}
\psi_{\mu - } &=- \frac{7\kappa^2}{20\lambda^2} e^{-\frac\sigma2}  H_{\mu \nu}^X \gamma^\nu \eta^X + \frac{3\kappa^2}{40\lambda^2} e^{-\frac\sigma2} H^{\rho \sigma X} \gamma_{\mu \rho \sigma} \eta^X + (\text{fermi})^3\ ,
\\
\chi & = \frac{\kappa^2}{20\lambda^2} e^{-\frac{\sigma}{2}} H_{\mu \nu}^X \gamma^{\mu\nu} \eta^X + (\text{fermi})^3\ ,
\\
 e^{6 \alpha \phi}\partial_\usevenn   \hat\sigma &=- \frac{\kappa^2}{10\lambda^2} e^{-\sigma} H_{\rho\sigma}^X H^{\rho \sigma X} + (\text{fermi})^2\ ,
 \\
G_{\mu \nu \usevenn} &= \frac{\kappa^2}{16\lambda^2} e^{-2 \sigma} \epsilon_{\mu\nu\rho\sigma\lambda\tau} H^{\rho \sigma X} H^{\lambda \tau X} + (\text{fermi})^2 \ ,
\\
K_{\mu\nu} &= \frac{\kappa^2}{2\lambda^2} e^{-\sigma} H_{\mu \rho}^X {H_\nu}^{\rho X} - \frac{3\kappa^2}{40\lambda^2} e^{-\sigma} H_{\rho\sigma}^X H^{\rho \sigma X} g_{\mu \nu}+ (\text{fermi})^2\ .
\end{split}
\label{BCtwo}
\end{equation}
Then, upon substituting this into the surface terms, obtained as before, a great deal of cancellation occurs and we are left with
\begin{equation}
\begin{split}
\delta_\epsilon S_{SG} + \delta_\epsilon S_{GHY} & = \frac{1}{\lambda^2} \int_{\partial M} d^6x e  \bigg \lbrace - \frac{1}{8} e^{ - \frac\sigma2 } \bar{\epsilon} \gamma^{\rho \sigma} \gamma^{\mu} \sigma^i \eta^X H_{\rho \sigma}^X \mathcal{P}_{\mu}^{i}  \bigg \rbrace\ .
\end{split}
\label{SurfVartwo}
\end{equation}
Finally, including a boundary action\footnote{The bulk contribution \eqref{SurfVartwo} can also be produced by adding a term of the form
\begin{equation}
S = \frac{1}{\lambda^2} \int_{\partial M} d^6 x e \bigg \lbrace \bar{\eta}^X \gamma^\mu \sigma^i \eta^X \mathcal{P}_\mu^i \bigg \rbrace
\end{equation}
to the boundary action and multiplying the R.H.S. of \eqref{BCtwo} by a corresponding factor. However, if this were done, the action and boundary conditions would then no longer be consistent with the variational principle.}
\begin{equation}
\begin{split}
S_B &=\frac{1}{\lambda^2} \int_{\partial M}  d^6x \bigg \lbrace - \frac18 e^{-\sigma}  H_{\mu \nu}^X H^{\mu \nu X}  - \frac{i}2 \bar{\eta}^X \gamma^\mu D_\mu \eta^X\\
& - \frac{i}4 e^{-\frac{\sigma}{2}} H_{\mu \nu}^X \bar{\eta}^X \gamma^\rho \gamma^{\mu \nu}  \psi_\rho -  \frac{i}4 e^{\frac{-\sigma}{2}} H_{\mu \nu}^X  \bar{\eta}^X \gamma^{\mu \nu} \chi
-\frac{i}{24}  e^{\sigma} G_{\mu \nu \rho} \bar{\eta}^X \gamma^{\mu \nu \rho} \eta^X
\\
& - \frac1{16} \epsilon^{\mu\nu\rho\sigma\lambda\tau} B_{\mu \nu} H_{\rho\sigma}^X H_{\lambda \tau}^X +
\frac1{32g^2} \epsilon^{\mu\nu\rho\sigma\lambda\tau} \omega^0_{\mu\nu\rho}(C) \omega^0_{\sigma\lambda\tau} (A)\bigg \rbrace\ ,
\end{split}
\end{equation}
gives the classical supersymmetry anomaly
\begin{equation}
\begin{split}
\delta_{\epsilon} S &= - \frac{1}{32\lambda^2 g^2} \int_{\partial M} d^6 x e \bigg \lbrace \epsilon^{\mu\nu\rho\sigma \lambda \tau} H_{\mu \nu}^X H_{\rho\sigma}^X \delta_\epsilon A_\lambda^{r^\prime} A_\tau^{r^\prime}
- 2 \epsilon^{\mu\nu\rho\sigma \lambda \tau} \omega_{\mu\nu\rho}^0 (C) F_{\sigma \lambda}^{r^\prime} \delta_\epsilon A_\tau^{r^\prime} \\
& + \epsilon^{\mu\nu\rho\sigma \lambda \tau} F_{\mu \nu}^{r^\prime} F_{\rho\sigma}^{r^\prime} \delta_\epsilon C_\lambda^X C_\tau^X
- 2 \epsilon^{\mu\nu\rho\sigma \lambda \tau} \omega_{\mu\nu\rho}^0 (A) H_{\sigma \lambda}^X \delta_\epsilon C_\tau^X \bigg \rbrace\ ,
\end{split}
\label{AnomClasSUSYtwo}
\end{equation}
whilst the classical gauge anomaly is
\begin{equation}
\delta_\Lambda S = - \frac1{32\lambda^2 g^2} \int_{\partial M} d^6 x e \bigg \lbrace  \varepsilon^{\mu \nu\rho \sigma \lambda \tau} H_{ \mu \nu}^X H_{\rho \sigma}^X \partial_\lambda A_\tau^{r^\prime} \Lambda^{r^\prime} + \varepsilon^{\mu \nu\rho \sigma \lambda \tau} F_{ \mu \nu}^{r^\prime} F_{\rho \sigma}^{r^\prime} \partial_\lambda C_\tau^{X} \Lambda^{X} \bigg \rbrace\ .
\label{AnomClasGaugetwo}
\end{equation}
Once again these are Wess-Zumino consistent.

It is interesting to note that these classical anomalies exist, in spite of the fact that there is no Chern-Simons term to provide anomaly inflow, because the inherited supergravity transformation rules have forced a Green-Schwarz type of anomaly production upon us. This is very different mechanism from the 3-form case considered in Section \ref{BoundaryAction}, but gives rise to anomalies of exactly the same form.

\end{appendices}

\newpage

\bibliography{HW7D}

\end{document}